\documentclass[aps,pra,notitlepage,twocolumn,10pt,a4paper]{revtex4-1}

\usepackage{xcolor,graphicx,ulem,soul}
\usepackage{amsmath,amssymb}
\usepackage[colorlinks=true,urlcolor=blue,citecolor=blue,linkcolor=blue]{hyperref}

\begin{document}

\title{Displaced photon-number entanglement tests}

\author{B. K\"uhn}\email{benjamin.kuehn2@uni-rostock.de}
\author{W. Vogel}
\affiliation{Arbeitsgruppe Quantenoptik, Institut f\"ur Physik, Universit\"at Rostock, D-18051 Rostock, Germany}

\author{J. Sperling}
\affiliation{Clarendon Laboratory, University of Oxford, Parks Road, Oxford OX1 3PU, United Kingdom}

\date{\today}

\begin{abstract}
        Based on correlations of coherently displaced photon numbers, we derive entanglement criteria for the purpose of verifying non-Gaussian entanglement.
	Our construction method enables us to verify bipartite and multipartite entanglement of complex states of light.
	An important advantage of our technique is that the certified entanglement persists even in the presence of arbitrarily high, constant losses.
	We exploit experimental correlation schemes for the two-mode and multimode scenarios, which allow us to directly measure the desired observables.
	To detect entanglement of a given state, a genetic algorithm is applied to optimize over the infinite set of our constructed witnesses.
	In particular, we provide suitable witnesses for several distinct two-mode states.
	Moreover, a mixed non-Gaussian four-mode state is shown to be entangled in all possible nontrivial partitions.
\end{abstract}

\maketitle


\section{Introduction}
	Quantum entanglement \cite{Einstein1935,Schroedinger1935} is an important resource for quantum technologies.
	For example, it serves as the basis for secure communication protocols \cite{Gisin2007} and quantum information processing \cite{Nielsen2000}.
	However, identifying entanglement of mixed, multipartite states remains a sophisticated task \cite{Horodecki2009,Guhne2009}.
	In particular, the quantum optical implementation of entangled light fields for applications requires novel methods to uncover entanglement in the distinct regimes of single-photon \cite{Walther2005} and continuous variables \cite{Ukai2011}.
	Recently, intermediate instances of hybrid systems at the interface of joint discrete- and continuous-variable quantum information have also gained a lot of attention; see, e.g., Ref. \cite{Andersen2015}.

	The most prominent approach to certifying entanglement is formulated in terms of so-called entanglement witnesses \cite{Horodecki1996,Horodecki2001}.
	Such observables have a limited range of expectation values attainable for separable states, and entanglement is accessed by violating those bounds \cite{Toth2005,Sperling2009}.
	In principle, this approach defines a necessary and sufficient method to characterize entanglement.
	Yet, the construction of witnesses is as challenging as the separability problem itself \cite{Gurvits2003}.
	Nonetheless, properly designed witnesses have been experimentally applied to successfully detect entanglement of certain classes of states; see, e.g., Ref. \cite{Bourennane2004} for an early implementation.

	Since a single (linear) entanglement witness cannot detect the entanglement of all quantum states, optimization procedures have been formulated \cite{Lewenstein2000}.
	For instance, the notion of finer or ultrafine witnesses \cite{Shahandeh2017} has been established to account for additional, physical constraints of the system under study.
	Also, the direct construction of optimal witnesses is a sophisticated problem which can be considered to ensure the best possible performance of entanglement tests for certain states.
	One consistent approach to such a desirable construction scheme is based on the method of separability eigenvalue equations \cite{Sperling2009,Sperling2013}.
	The solution of those equations allows for the formulation of optimized entanglement criteria which, in principle, render it possible to derive witnesses for arbitrary detection schemes.
	This versatile approach has been used to experimentally characterize path-entangled photons in the single-photon domain \cite{Gutierrez2014} and to uncover complex forms of multimode entanglement in continuous-variable Gaussian states of light \cite{Gerke2015,Gerke2016}.

	Of major importance is the class of entangled non-Gaussian states, which plays a crucial role in several quantum applications.
	In particular, these states are necessarily required in certain protocols for entanglement distillation \cite{Eisert2002}, quantum error correction \cite{Niset2009}, and quantum teleportation \cite{Seshadreesan2015}.
	Therefore, it is essential to have powerful entanglement criteria, which are based on higher-than-second-order moments and which are able to certify the entanglement of non-Gaussian states. 
	Some criteria are proposed in Refs. \cite{Shchukin2005,Gomes2009,Hertz2016}.
	
	As indicated above, the availability of entanglement not only depends on the sources of multimode-correlated quantum light, but also relies on the availability of detection schemes which allow for certification of entanglement.
	For example, the measurement of multimode photon-number correlations is not sufficient to infer entanglement.
	Phase-randomized states, such as those constructed in Refs. \cite{Ferraro2012} and \cite{Agudelo2013}, can exhibit the same form of photon-number correlations, and they are separable at the same time.
	Moreover, experimental techniques in discrete- and continuous-variable quantum optics require rather distinct resources.
	
	A standard technique to verify continuous-variable entanglement relies on multimode balanced homodyne detection for the measurement of the covariance matrix of Gaussian states \cite{Roslund2014,Medeiros2014,Chen2014}. 
	This provides all necessary information, e.g., for the Simon \cite{Simon2000} or the Duan \textit{et al.} \cite{Duan2000} entanglement criteria, to test entanglement of all the bipartitions. 
	Beyond bipartitions, from the covariance matrix one can even verify the entanglement of all individual multipartitions of Gaussian states~\cite{Gerke2015,Gerke2016}. 
	For the more general task of analyzing multipartite non-Gaussian entanglement, it is important to merge the theoretical construction of entanglement tests with the availability of proper measurement techniques to implement promising test strategies.

	In this work, we derive a class of entanglement witnesses which apply to any combination of displaced photon-number correlations.
	We use our approach to study entanglement of bipartite and multipartite radiation fields.
	Experimental methods are considered which are suited to directly access our criteria in the discrete- and continuous-variable regime.
	Furthermore, we observe that our displaced photon-number witnesses are robust against constant losses.

	Our work is organized as follows.
	In Sec. \ref{ch:DPNW}, entanglement witnesses on the basis of displaced photon-number correlations are constructed for the bipartite case and their properties are analyzed.
	An experimental scheme is proposed in Sec. \ref{ch:ExpImp}, which enables direct measurement of such witnesses, and relations to other notions of quantum correlations and the robustness of our method are discussed.
	In Sec. \ref{ch:Application}, we characterize the entanglement of several, relevant quantum states.
	We generalize our treatment to the multimode scenario in Sec. \ref{ch:Multimode}.
	In Sec. \ref{ch:Conclusions}, we summarize and conclude.

\section{Displaced photon-number witnesses for bipartite entanglement}\label{ch:DPNW}
	In this section, first we briefly recall the method used to construct optimal entanglement witnesses~\cite{Sperling2009}. 
	Then we consider some properties of the displaced photon-number operator. 
	Eventually, we formulate our entanglement criteria---based on displaced photon-number measurements---for the bipartite scenario.
	
	In the context of this section, let us also recall the notions of bipartite separability \cite{Werner89}.
	Namely, a pure state $|\psi\rangle$ in the compound Hilbert space $\mathcal H_{a}\otimes\mathcal H_{b}$ is separable by definition if it is a normalized tensor product, $|\psi\rangle=|a\rangle\otimes|b\rangle$.
	By extension, a mixed state $\hat\sigma$ is separable if it is a statistical mixture of pure separable states,
	\begin{align}\label{eq:SepStates}
		\hat\sigma=\int dP(a,b)|a\rangle\langle a|\otimes|b\rangle\langle b|,
	\end{align}
	where $P$ is a probability distribution over the set of pure separable states.
	Any state which cannot be expanded in this form is entangled.

\subsection{Entanglement witnesses from separability eigenvalue equations}
	The construction of entanglement witnesses can be done, e.g., through the optimization of the expectation value of a given Hermitian operator $\hat L$ with respect to all separable states \cite{Toth2005,Sperling2009}.
	Due to convexity, it is sufficient to consider the optimization over pure separable states $|x\rangle\otimes|y\rangle$.
	Depending on the considered scenario, either a maximization or a minimization can be more advantageous.
	Here, we focus on the minimization procedure.
	Note that the maximization of expectation values of $\hat L$ is identical to the minimization of those of $-\hat L$.

	The considered minimization leads to a problem \cite{Sperling2009}, which is defined in terms of the so-called separability eigenvalue equations (SEEs),
	\begin{subequations}
	\begin{eqnarray}\label{eq:SEPintro}
		\hat L_y|x\rangle&=&g|x\rangle\label{eq:SEPintroa},\\
		\hat L_x|y\rangle&=&g|y\rangle\label{eq:SEPintrob},
	\end{eqnarray}
	\end{subequations}
	which include reduced operators defined as
	\begin{subequations}
	\begin{eqnarray}\label{eq:ReducedOpA}
		\hat L_x&=&\mathrm{tr}_a[\hat L(|x\rangle\langle x|\otimes\hat 1)],
		\\\label{eq:ReducedOpB}
		\hat L_y&=&\mathrm{tr}_b[\hat L(\hat 1\otimes|y\rangle\langle y|)],
	\end{eqnarray}
	\end{subequations}
	using the normalized vectors $|x\rangle\in \mathcal{H}_{a}$ and $|y\rangle\in \mathcal{H}_{b}$.
	The real number $g$ denotes the separability eigenvalue (SEV) of the Hermitian operator $\hat L$, and the state $|x\rangle\otimes |y\rangle$ denotes the corresponding separability eigenstate (SES).
	Note that Eqs. \eqref{eq:SEPintroa} and \eqref{eq:SEPintrob} are coupled since the solution vector $|y\rangle$ of the second equation defines the reduced operator $\hat L_{y}$ for the first equation, and vice versa.

	Using this technique, it has been shown that one can formulate separability constraints \cite{Sperling2009}.
	That is, for any separable state it holds that
	\begin{align}\label{eq:EntCriterion}
		\langle \hat L\rangle\geq g_{\min}=\inf\{g: \text{SEV to }\hat L\}.
	\end{align}
	A violation of this bound, $\langle \hat L\rangle<g_{\min}$, identifies entanglement.
	Hence, the minimal SEV allows for the formulation of entanglement criteria.
	Furthermore, it is worth mentioning that for a rescaled and shifted operator, $\mu\hat L+\nu\hat 1\otimes\hat 1$ for $\mu,\nu\in\mathbb R$ and $\mu>0$, the minimal SEV reads $\mu g_{\min}+\nu$ \cite{Sperling2013}.
	In addition, an entanglement witness can be directly obtained from constraint \eqref{eq:EntCriterion} and $\langle \hat 1\otimes\hat 1\rangle=1$,
	\begin{align}\label{eq:generalwitness}
		\hat W=\hat L-g_{\min}\hat 1\otimes\hat 1,
	\end{align}
	which has nonnegative expectation values for all separable states.

	The generalization of this technique to an $N$-partite system yields SEEs containing $N$ coupled equations \cite{Sperling2013}, which are applied in Sec.~\ref{ch:Multimode}.
	Using this multimode generalization, an experimentally generated $10$-mode frequency-comb Gaussian state was successfully tested for entanglement in all possible nontrivial partitions \cite{Gerke2015}.
	Also, the concept of partial entanglement itself was further extended and experimentally applied to convex combinations of individual partitions \cite{Gerke2016}.

\subsection{Single-mode eigenvalue problem of displaced photon-number observables}
	As we study displaced photon-number statistics, let us recall some of their properties; see Ref. \cite{Vogel2006} for an introduction.
	One bosonic mode is represented through the annihilation (creation) operator $\hat a$ ($\hat a^\dag$).
	The photon-number operator is $\hat n=\hat a^\dag\hat a$.
	Furthermore, the unitary displacement operator, $\hat D(\alpha)=\exp[\alpha\hat a^\dag-\alpha^\ast\hat a]$ for $\alpha\in\mathbb C$, allows one to define the displaced photon-number operator as
	\begin{align}
		\hat n(\alpha)=\hat D(\alpha)\hat n\hat D(\alpha)^\dag=(\hat a-\alpha)^\dag(\hat a-\alpha).
	\end{align}
	Because of the unitary transformation, this operator has the eigenvalues $n\in\mathbb N$ and eigenstates which are displaced photon-number states, $\hat D(\alpha)|n\rangle$.
	In particular, the ground state ($n=0$) yields the coherent states of the quantized radiation field, $\hat D(\alpha)|0\rangle=|\alpha\rangle$.

	Let us now consider a combination of displaced photon-number operators,
	\begin{align}\label{eq:SingleModeOp}
		\hat L=\sum_{k=1}^m \lambda_k \hat n(\alpha_k),
	\end{align}
	where $\lambda_k>0$ and $\alpha_k\neq\alpha_{k'}$ for all $k\neq k'$.
	Because of the scaling properties of the (separability) eigenvalue equations \cite{Sperling2009,Sperling2013}, we can additionally assume that $\sum_k\lambda_k=1$.
	Therefore, we can interpret $\{\lambda_k\}_{k=1,\ldots,m}$ as a probability distribution over the random variable $A$ of coherent amplitudes $\{\alpha_k\}_{k=1,\ldots,m}$.
	For example, the mean coherent amplitude is given by $\overline{A}=\sum_{k}\lambda_k \alpha_k$.

	Rewriting the operator, \eqref{eq:SingleModeOp}, yields a combination of a displaced photon-number operator and the identity,
	\begin{align}
		\hat L=\hat n(\overline A)+\overline{|\Delta A|^2}\hat 1,
	\end{align}
	where $\Delta A=A-\overline{A}$.
	The minimal eigenvalue of this operator is $\overline{|\Delta A|^2}$, which is attained for a coherent state $|\alpha\rangle$ with $\alpha=\overline A$.

\subsection{Displaced photon-number correlations}\label{subsec:DPNC2mode}
	In the next step, let us combine the previously discussed relations to formulate bipartite entanglement criteria.
	Similarly to the single-mode operator, \eqref{eq:SingleModeOp}, we consider a test operator for the two-mode system of the form
	\begin{eqnarray}\label{eq:testoperator}
		\hat L=\sum_{k=1}^{m}\lambda_k\hat n(\alpha_k)\otimes\hat n(\beta_k), 
	\end{eqnarray}
	for pairwise different complex displacements, $(\alpha_k,\beta_k)\neq(\alpha_{k'},\beta_{k'})$ for all $k\neq {k'}$.
	The positive weighting factors $\{\lambda_k\}_{k=1,\dots,m}$ are normalized to guarantee that $\sum_k\lambda_k=1$.
	The expectation value of the operator in Eq. \eqref{eq:testoperator} is a convex combination of $m$ two-mode displaced photon-number correlations, $\langle \hat n(\alpha_k)\otimes\hat n(\beta_k)\rangle$.
	Moreover, this operator is positive semidefinite, and its expectation value is unbounded from above;
	for example, we have $\langle\hat L\rangle\to\infty$ for states $|\alpha\rangle\otimes|\beta\rangle$ with amplitudes $|\alpha|\to\infty$ or $|\beta|\to\infty$.
	It is also important to point out that $\hat L$ is intrinsically a non-Gaussian operator since it contains up to fourth-order terms of the annihilation and creation operators.

	To get entanglement criteria, we have to solve the SEEs in Eqs. \eqref{eq:SEPintroa} and \eqref{eq:SEPintrob}.
	This requires to compute the reduced operators in Eqs. \eqref{eq:ReducedOpA} and \eqref{eq:ReducedOpB}, which read
	\begin{subequations}
	\begin{eqnarray}
		\hat L_{x}&=&\sum_{k=1}^m\lambda_k\langle {x}|\hat n(\alpha_k)|{x}\rangle\hat n(\beta_k),
		\\
		\hat L_{y}&=&\sum_{k=1}^m\lambda_k\langle {y}|\hat n(\beta_k)|{y}\rangle\hat n(\alpha_k).
	\end{eqnarray}
	\end{subequations}
	They are both of the form of \eqref{eq:SingleModeOp}.
	Therefore, the eigenstate to the minimal eigenvalue of both reduced operators has the form of a coherent state.
	This allows us to conclude that the SES to the minimal SEV is a product of coherent states, $|{x}\rangle\otimes|{y}\rangle=|\alpha\rangle\otimes|\beta\rangle$.

	Applying these considerations to the SEEs, \eqref{eq:SEPintroa} and \eqref{eq:SEPintrob}, results in the following coupled equations for the complex amplitudes:
	\begin{subequations}
	\begin{align}
		\sum_{k=1}^m\lambda_k|\beta{-}\beta_k|^2(\alpha{-}\alpha_k)
		=\overline{|\beta-B|^2(\alpha-A)}
		&=0,\label{eq:coupleda}\\
		\sum_{k=1}^m\lambda_k|\alpha{-}\alpha_k|^2(\beta{-}\beta_k)
		=\overline{|\alpha-A|^2(\beta-B)}
		&=0\label{eq:coupledb},
	\end{align}
	\end{subequations}
	where we have used the interpretation of $\{\lambda_k\}_{k=1,\ldots,m}$ as a probability distribution for the pair of random variables $(A,B)$ with values $\{(\alpha_k,\beta_k)\}_{k=1,\ldots,m}$.
	In addition, the minimal SEV is given by
	\begin{eqnarray}\label{eq:minSEVbipart}
		&&g_{\min}=\overline{|\alpha-A|^2|\beta-B|^2}.
	\end{eqnarray}
	In total, the test operator $\hat L$ depends on $4m+(m-1)$ independent, real-valued parameters.
	These are the positive numbers $\lambda_k$ (minus normalization) and the real and imaginary parts of the displacements $\alpha_k$ and $\beta_k$.

	In addition, let us stress that the number, \eqref{eq:minSEVbipart}, is the desired bound of the separability constraint, \eqref{eq:EntCriterion}, for the given observable, \eqref{eq:testoperator}.
	Also, note that the values of $\alpha$ and $\beta$ are not explicitly determined.
	Still, they can be obtained as the roots of polynomials; cf. Eqs. \eqref{eq:coupleda} and \eqref{eq:coupledb}.
	It is also worth mentioning that a local displacement, $[\hat D(\alpha_s)\otimes\hat D(\beta_s)]\hat L [\hat D(\alpha_s)\otimes\hat D(\beta_s)]^\dagger$, of the operator $\hat L$ in both modes does not change the value of the minimal SEV.
	Rather it transforms the SESs to $|\alpha+\alpha_s\rangle\otimes|\beta+\beta_s\rangle$.

\subsection{Special cases}\label{subsec:SpecialCases}
	Let us now take a closer look at the number of terms $m$, which define the test operator, \eqref{eq:testoperator}.
	For $m=1$, we get the SES $|\alpha_1\rangle\otimes|\beta_1\rangle$, which results in the SEV $g_{\min}=0$.
	In the case $m=2$, we get $g_{\min}=0$ for the SESs $|\alpha_1\rangle\otimes|\beta_2\rangle$ and $|\alpha_2\rangle\otimes|\beta_1\rangle$.
	Since $\hat L$ is positive semidefinite, i.e., $\langle \hat L\rangle\geq0$ for any state, constraint \eqref{eq:EntCriterion} is always satisfied in this scenario.
	Therefore, $m\geq3$ is required to be able to certify entanglement.

	Now, let us restrict ourselves to the case $m=3$ and to displacements that lie on a line in phase space, i.e.,  $\alpha_k=\underline{\alpha}_ke^{i\phi},\,\beta_k=\underline{\beta}_ke^{i\theta}$, with fixed phases $\phi$ and $\theta$ and $\underline{\alpha}_k,\underline{\beta}_k\in\mathbb R$.
	Solving Eq. \eqref{eq:coupleda} for $\alpha$ and Eq. \eqref{eq:coupledb} for $\beta$, one obtains the phase relations $\alpha=\underline{\alpha}e^{i\phi}$ and $\beta=\underline{\beta}e^{i\theta}$ with real-valued parameters $\underline{\alpha},\underline{\beta}$.
	For these amplitudes, we get
	\begin{subequations}
	\begin{eqnarray}
		\underline{\alpha}&=&\dfrac{\sum_{k=1}^3\lambda_k(\underline{\beta}-\underline{\beta}_k)^2\underline{\alpha}_k}{\sum_{k=1}^3\lambda_k(\underline{\beta}-\underline{\beta}_k)^2},
		\label{eq:a}\\
		\underline{\beta}&=&\dfrac{\sum_{k=1}^3\lambda_k(\underline{\alpha}-\underline{\alpha}_k)^2\underline{\beta}_k}{\sum_{k=1}^3\lambda_k(\underline{\alpha}-\underline{\alpha}_k)^2}.
		\label{eq:b}
	\end{eqnarray}
	\end{subequations}
	Inserting Eq. \eqref{eq:a} into Eq. \eqref{eq:b}, one obtains a root finding problem for a polynomial with a degree of 5,
	\begin{eqnarray}\label{eq:poly}
		\sum_{k=0}^5 c_k\underline{\beta}^k=0,
	\end{eqnarray}
	with the coefficients
	\begin{align}
	\begin{aligned}
		c_0&=-\sum_{k=1}^3\lambda_k R_{2k}^2\underline{\beta}_k,\\
		c_1&=\sum_{k=1}^3\lambda_k R_{2k}(R_{2k}+4R_{1k}\underline{\beta}_k),\\
		c_2&=-2\sum_{k=1}^3\lambda_k \left[(2 R_{1k}^2+R_{2k}R_{0k})\underline{\beta}_k+2R_{1k}R_{2k}\right],\\
		c_3&=2\sum_{k=1}^3\lambda_k (2R_{1k}^2+R_{2k}R_{0k}+2R_{1k}R_{0k}\underline{\beta}_k),\\
		c_4&=-\sum_{k=1}^3\lambda_k R_{0k}(R_{0k}\underline{\beta}_k+4R_{1k}),\\
		c_5&=\sum_{k=1}^3\lambda_k R_{0k}^2,
	\end{aligned}
	\end{align}
	and
	\begin{eqnarray}
		R_{\ell j}&=&\sum_{k=1}^3\lambda_k(\underline{\alpha}_k-\underline{\alpha}_j)\underline{\beta}_k^\ell,
	\end{eqnarray}
	for $\ell=0,1,2$ and $j=1,2,3$.
	Equation \eqref{eq:poly} can be solved numerically, and $\underline{\alpha}$ can be inferred from Eq. \eqref{eq:a}.

\section{Relations, implementation, and imperfections}\label{ch:ExpImp}
	In the previous section, we formulated a technique to infer entanglement in terms of criteria which are based on displaced photon-number correlations.
	In this section, we relate the verified entanglement to other notions of correlations in the first step.
	Then, we propose an experimental technique to measure the expectation value $\langle\hat L\rangle$ of the operator in Eq. \eqref{eq:testoperator}.
	It is based on a two-mode correlation measurement, and it can be straightforwardly generalized to multimode scenarios.
	Finally, the robustness of our approach under attenuations is elaborated.

\subsection{Relation to other notions of quantumness}
	Beyond entanglement, there are also other quantum correlations in radiation fields; see, e.g., Refs. \cite{Ferraro2012}, \cite{Agudelo2013}, and \cite{Vo08}.
	In particular, in quantum optics, nonclassicality of a two-mode radiation field is defined on the basis of a Glauber-Sudarshan $P$ function \cite{Glauber1963,Sudarshan1963,Titulaer1965}.
	All two-mode quantum states can be represented via this function in terms of coherent states as
	\begin{align}\label{eq:GSrepresentation}
		\hat\rho=\int d^2\alpha \,d^2\beta\,P(\alpha,\beta)\,|\alpha\rangle\langle\alpha|\otimes|\beta\rangle\langle\beta|.
	\end{align}
	If $P$ cannot be interpreted in terms of a classical probability distribution, the state is referred to as nonclassical.
	That is, a state exhibits nonclassical correlations, and it cannot be considered as a convex mixture of coherent states $|\alpha\rangle\otimes|\beta\rangle$.

	Since the $P$ function can be highly singular, which may prevent a direct certification of quantumness, one can use so-called nonclassicality witnesses, whose expectation value is nonnegative for all classical states; see Ref. \cite{Sperling2016} for a recent study.
	In close analogy to entanglement witnesses, a nonclassicality witness can be constructed by starting with a Hermitian operator and optimizing its expectation value with respect to the convex set of classical states.
	A general comparison of nonclassicality and entanglement criteria can be found in Ref. \cite{Miranowicz2010}.
	For our observable in terms of displaced photon-number correlations [Eq. \eqref{eq:testoperator}], we already know that the minimal expectation value for separable states is attained for a tensor product of coherent states.
	Thus, the minimal expectation value for separable and classically correlated states is identical for both scenarios.
	Hence, verification of nonclassical correlations in the sense of the standard notion in quantum optics also implies entanglement. 

	This is consistent with the finding in Ref. \cite{Sperling2009P2}.
	There it was shown that optimal entanglement quasiprobability distributions can be computed by solving the SEEs for the density operator.
	This entanglement quasiprobability can include negative contributions, and it requires that the Glauber-Sudarshan $P$ function be a nonclassical distribution as well.
	It further relates the definitions of separable states in Eq. \eqref{eq:SepStates} to the notion of nonclassicality based on Eq. \eqref{eq:GSrepresentation}.
	That is, the restriction to coherent states, $|\alpha\rangle\otimes|\beta\rangle$, for the concept of nonclassicality is replaced by arbitrary product states, $|a\rangle\otimes|b\rangle$, to characterize entanglement. 
	As a consequence, a classically correlated state is also separable, but not the other way around.

\subsection{Measurement scheme}
	The photoelectric detection of light together with the interference of quantum light with coherent light is one way to access the displaced photon number \cite{Vogel2006}.
	Recently, multiplexing schemes together with imperfect detectors have been used to estimate quantum light with comparably high photon numbers; see, e.g., Ref. \cite{Harder2015}.
	Using such recent implementations, the nonclassicality of quantum light, in the sense of the $P$ function, has been inferred in a detector-independent manner \cite{Sperling2017}.
	Moreover, phase-sensitive measurements have been performed using such multiplexing detectors, e.g., in Ref. \cite{Donati2014}.
	In combination, these measurement strategies allow for the detection of nonclassical light in the regime between continuous-variable and single-photon quantum optics.
	Beyond that, we proposed a method to directly measure the displaced photon-number statistics using unbalanced homodyne correlation measurements \cite{Kuehn2016}.

\begin{figure}[ht]
	\includegraphics[clip,scale=0.65]{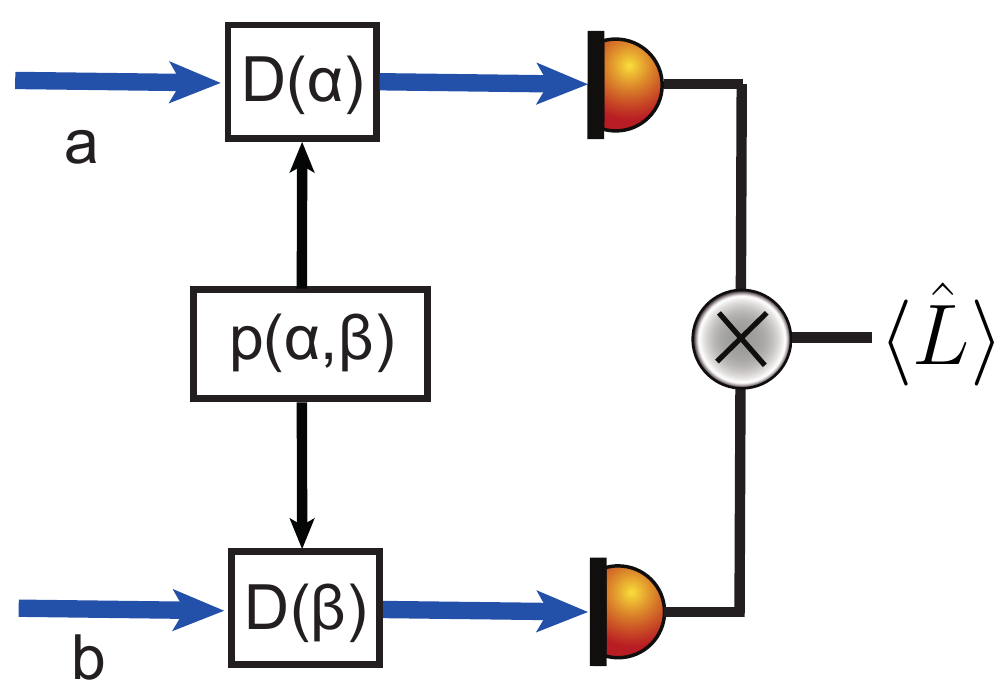}
	\caption{(Color online)
		Outline of an experimental scheme to measure $\langle\hat L\rangle$.
		The two modes of the incident light field, labeled ``a'' and ``b'', are displaced, which is indicated by the displacement operators $\hat D(\alpha)$ and $\hat D(\beta)$.
		The desired coherent displacements $\{(\alpha_k,\beta_k)\}_{k=1,\dots,m}$ are randomly realized according to the distribution $p(\alpha,\beta)$, which is defined through the discrete probabilities $\{\lambda_k\}_{k=1,\ldots,m}$.
	}\label{fig:DNESetup}
\end{figure}

	Here, a technique---compatible with such detection schemes---is constructed to infer entanglement, rather than nonclassicality.
	Figure \ref{fig:DNESetup} shows an outline of the thought experimental setup to detect entanglement.
	Modes $a$ and $b$ are coherently displaced by amplitudes $\alpha_k$ and $\beta_k$, respectively.
	This can be realized by superimposing each mode with a weak local oscillator on a highly transmitting beam splitter.
	Afterwards, the photon-number correlation can be measured by applying photon-number resolving detectors \cite{Divochiy2008,Jahanmirinejad2012,Smith2012,Calkins2013}. 
	Alternatively, linear detectors without photon-number resolution can be used, based on our unbalanced homodyne correlation measurement technique \cite{Kuehn2016}.
	For each measurement run, a (classical) random generator produces the displacements $\{(\alpha_k,\beta_k)\}_{k=1,\dots,m}$ according to the probabilities $\{\lambda_k\}_{k=1,\dots,m}$.
	In this manner, $\langle\hat L\rangle$ can be directly measured without the need for additional reconstruction algorithms or data postprocessing.

\subsection{Loss robustness}
	It is well known that experimental imperfections play a crucial role in the verification of quantum correlations.
	Especially, losses can affect the entanglement certification; see, e.g., Refs. \cite{Filippov2014} and \cite{Bohmann2016}.
	Here we consider finite detection efficiencies $\eta_{a}$ and $\eta_{b}$ ($\eta_a\eta_b\neq0$) of the two employed detectors (Fig. \ref{fig:DNESetup}).
	This leads to the transformed field operators $\hat a\mapsto\sqrt{\eta_a}\hat a$ and $\hat b\mapsto\sqrt{\eta_b}\hat b$, which result in a transformation of the operator $\hat L$ in Eq. \eqref{eq:testoperator} to
	\begin{align}\label{eq:LossTestOp}
		\hat L^{(\eta_{a},\eta_{b})}=\eta_{a}\eta_{b}\sum_{k=1}^m \lambda_k \hat n\left(\frac{\alpha_k}{\sqrt{\eta_a}}\right)\otimes\hat n\left(\frac{\beta_k}{\sqrt{\eta_b}}\right).
	\end{align}
	We find that the minimal SEV $g_{\min}$ of the operator $\hat L$ in Eq. \eqref{eq:testoperator} coincides with the minimal SEV $g_{\min}^{(\eta_a,\eta_b)}$ of the operator $\hat L^{(\eta_{a},\eta_{b})}$ in Eq. \eqref{eq:LossTestOp},
	\begin{align}\label{eq:LossSEV}
		g_{\min}^{(\eta_a,\eta_b)}=g_{\min};
	\end{align}
	see Appendix \ref{ch:appconstantloss} for the detailed derivation.
	
	Let us study the following situation.
	Assume that for a given state in the unperturbed scenario there exists a set of specific parameters $\{(\lambda_k,\alpha_k,\beta_k)\}_{k=1,\dots,m}$ for a test operator of the form of \eqref{eq:testoperator} such that entanglement is verified, $\langle\hat W\rangle<0$, with the witness operator 
	\begin{align}
		\hat W=\sum_{k=1}^m\lambda_k\hat n(\alpha_k)\otimes\hat n(\beta_k)-g_{\mathrm{min}}\hat 1\otimes\hat 1;
	\end{align}
	cf. Eq. \eqref{eq:generalwitness}.
	It is now interesting to investigate whether there exist parameters $\{(\tilde\lambda_k,\tilde\alpha_k,\tilde\beta_k)\}_{k=1,\dots,m}$ such that the witness, 
	\begin{align}
		&\hat W^{(\eta_{a},\eta_{b})}\notag\\
		&=\eta_a\eta_b\sum_{k=1}^m\tilde\lambda_k\hat n\left(\dfrac{\tilde\alpha_k}{\sqrt{\eta_a}}\right)\otimes\hat n\left(\dfrac{\tilde\beta_k}{\sqrt{\eta_b}}\right)-\tilde g^{(\eta_{a},\eta_{b})}_{\mathrm{min}}\hat 1\otimes\hat 1,
	\end{align}
	which is constructed from the transformed operator in Eq. \eqref{eq:LossTestOp}, certifies the entanglement of the state including the detection losses.
	Using Eq. \eqref{eq:LossSEV}, we can rewrite the expectation value of this witness as
	\begin{align}
		&\langle\hat W^{(\eta_{a},\eta_{b})}\rangle\notag\\
		=&\eta_a\eta_b
			\sum_{k=1}^m \tilde\lambda_k \left\langle
				\hat n\left(\frac{\tilde\alpha_k}{\sqrt{\eta_a}}\right)\otimes\hat n\left(\frac{\tilde\beta_k}{\sqrt{\eta_b}}\right)
			\right\rangle
			-\tilde g_{\min}.
	\end{align}
	Choosing $\tilde\lambda_k=\lambda_k$, as well as $\tilde\alpha_k=\sqrt{\eta_a}\alpha_k$, and $\tilde\beta_k=\sqrt{\eta_b}\beta_k$, it follows that
	\begin{align}
		\tilde g_{\min}=\eta_a\eta_b\,g_{\min},
	\end{align}
	and, thus, we directly observe that
	\begin{align}
		\langle\hat W^{(\eta_{a},\eta_{b})}\rangle=\eta_a\eta_b\langle\hat W\rangle<0.
	\end{align}
	This means that using modified displacements for lossy detection scenarios, with $\eta_a\eta_b\neq0$, we can still detect entanglement.
	This is an important finding since it means that arbitrarily high, constant losses do not annihilate the detectable entanglement--or, in other words, the non-Gaussian entanglement, certified through displaced photon-number correlations, cannot be destroyed by detection losses.

\section{Application}\label{ch:Application}
	In this section, we apply our technique to uncover entanglement of various two-mode states.
	For instance, we will provide a method to find the optimal parameters $\{(\lambda_k,\alpha_k,\beta_k)\}_{k=1,\dots,m}$ of the observable, \eqref{eq:testoperator}, for certifying entanglement of a particular state.
	Our examples include the non-Gaussian Schr\"odinger-cat-like states, two-mode Gaussian states, and non-Gaussian states produced via photon subtraction.

	Before we consider the individual families of states, let us briefly outline the general application of the method introduced in Sec. \ref{ch:DPNW}.
	To apply the operator $\hat L$ [Eq. \eqref{eq:testoperator}] to a given state, we need to identify suitable configurations of the parameters $\lambda_k$, $\alpha_k$, and $\beta_k$.
	In particular, we focus on three possible displacement configurations, $m=3$; see Sec. \ref{subsec:SpecialCases}.
	Moreover, it turns out that for our examples, the weighting factors $\{\lambda_1,\lambda_2,\lambda_3\}$ can be chosen to be equal, $\lambda_k=1/3$.

	Hence, we have to determine the coherent displacements $\{\alpha_1,\alpha_2,\alpha_3\}$ and $\{\beta_1,\beta_2,\beta_3\}$ for an optimal verification of entanglement.
	This is done by implementing a so-called genetic optimization; see Ref. \cite{Haupt2004} for an introduction.
	This approach was also previously applied to find optimal, Gaussian entanglement tests \cite{Gerke2015,Gerke2016}.
	In our case, the underlying algorithm minimizes the expression $\langle\hat L\rangle-g_{\mathrm{min}}$ over the possible coherent displacements and, thereby, determines the coherent displacements.
	It is also worth emphasizing that it is important to reach a maximal violation of the separability constraint, \eqref{eq:EntCriterion}, as well as to choose the parameters in an experimentally simple way, e.g., with a high degree of symmetry.

\subsection{Two-mode superposition of coherent states}
	Let us start by considering the non-Gaussian state
	\begin{align}\label{eq:BellState}
		|\psi\rangle=\mathcal{N}\left[\left(1-\dfrac{|\epsilon|}{2}\right)|\gamma\rangle\otimes|-\gamma\rangle+\dfrac{\epsilon}{2}|-\gamma\rangle\otimes|\gamma\rangle\right],
	\end{align}
	with coherent states $|\gamma\rangle$ and $|-\gamma\rangle$, normalization constant $\mathcal{N}$, and complex parameter $\epsilon$.
	For $|\epsilon|=1$, we have a balanced superposition of the two separable contributions.
	Since this state superimposes two linearly independent product states, it can be related to a Schr\"odinger cat state.
	Likewise, it is a continuous-variable analog to a Bell state.
	In the following, we restrict ourselves to a coherent amplitude of $\gamma=0.6$.

\begin{figure}[ht]
	(a)\\
	\includegraphics[clip,scale=0.95]{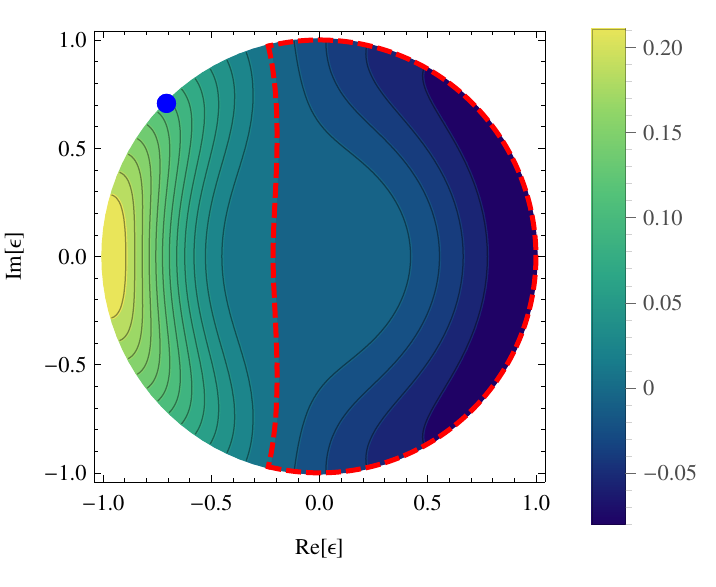}
	\\(b)\\
	\includegraphics[clip,scale=0.95]{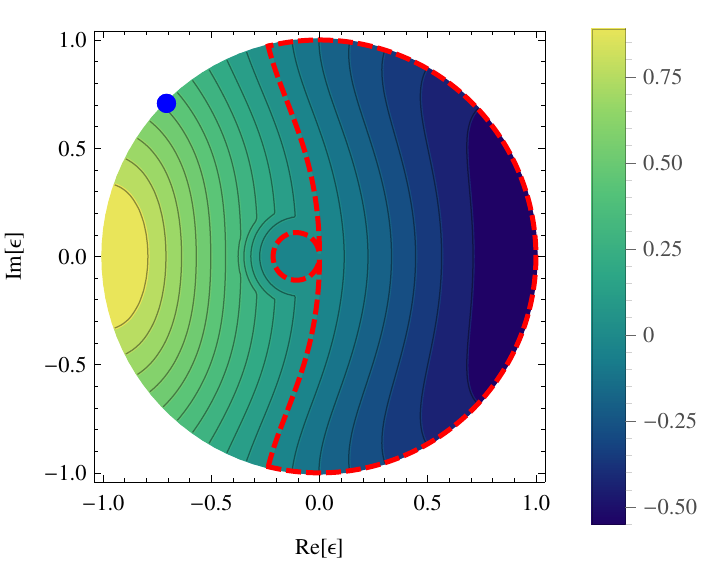}
	\caption{(Color online) 
		The inseparability criteria by (a) Simon and (b) Duan \textit{et al.} for the state \eqref{eq:BellState}, with $\epsilon\in\mathbb C$ and $0<|\epsilon|\leq 1$.
		The negative region, bounded by the dashed red curve, successfully probes the entanglement.
		For nonnegative values, such as for the example $\epsilon=e^{i(3/4)\pi}$ (blue point), no entanglement is identified by those criteria.
	}\label{fig:simonduan}
\end{figure}

	We first use two typically applied, covariance-based entanglement criteria in continuous variables.
	In Fig. \ref{fig:simonduan}, the criteria proposed by Simon \cite{Simon2000} and Duan \textit{et al.} \cite{Duan2000} are depicted for $|\epsilon|\leq 1$.
	It is worth mentioning that both approaches are based on the partial transposition \cite{Peres1996} and that both criteria are closely related to each other \cite{Horodecki2009}.
	Entanglement is uncovered for negative values in Fig. \ref{fig:simonduan}.
	For example, both criteria fail to demonstrate the entanglement of state \eqref{eq:BellState} for $\epsilon=e^{i(3/4)\pi}$ (angle of $135^\circ$; blue circle).

\begin{figure}[ht]
	\includegraphics[clip,scale=0.55]{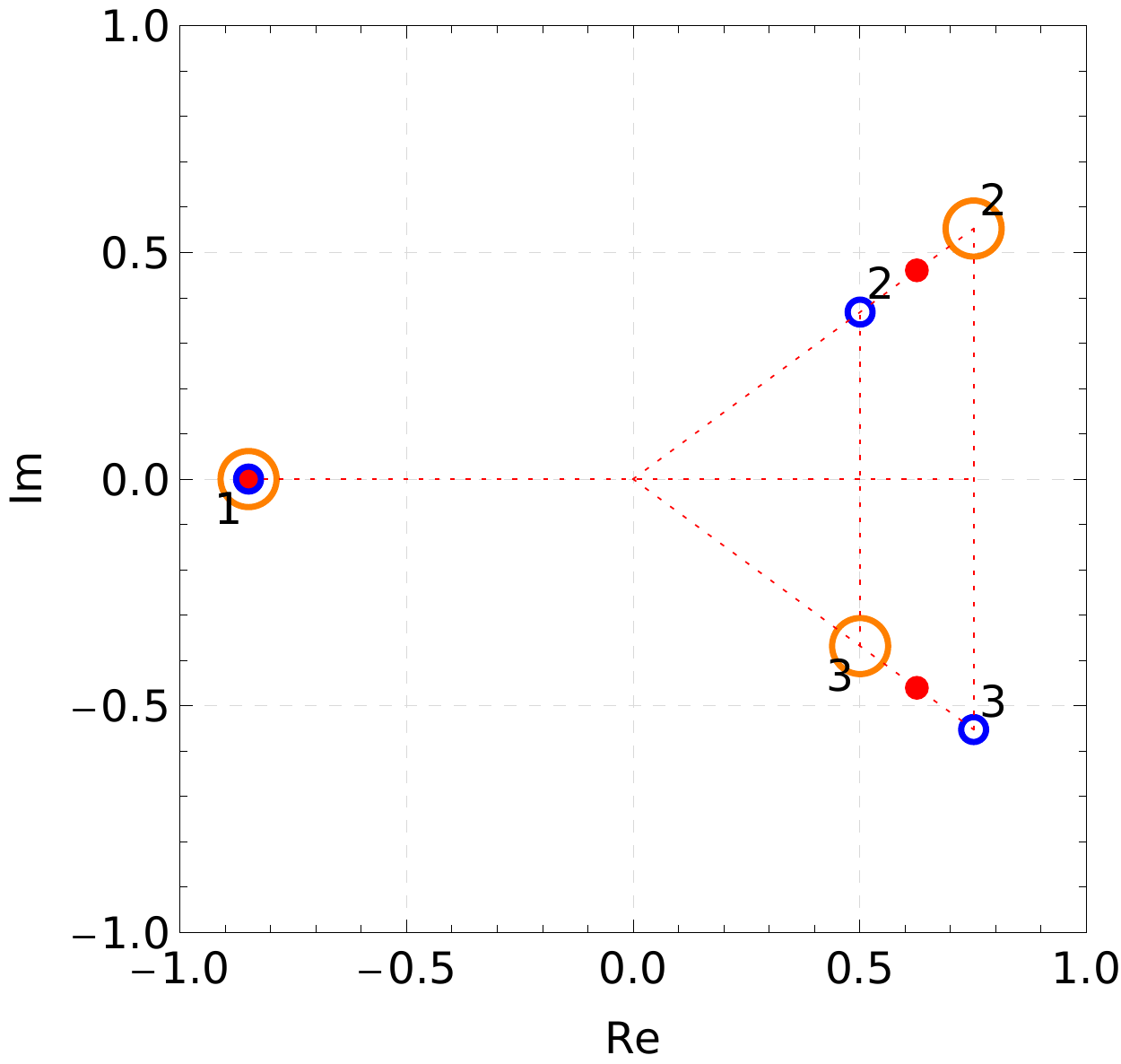}
	\caption{(Color online) 
		Displacements $\alpha_k$ (larger, orange circles) and $\beta_k$ (smaller, blue circles) with $k=1,2,3$ for witnessing entanglement of state \eqref{eq:BellState} are shown in the complex plane. Re and Im denote the real and imaginary axes, respectively.
		Small filled red circles depict $\left\{Q_1,Q_2,Q_3\right\}$.
	}\label{fig:confSupCoh}
\end{figure}

	Let us apply our approach to this particular value, $\epsilon=e^{i(3/4)\pi}$.
	Our method yields the coherent amplitudes---discussed in the next paragraph---for which an optimal difference between the expectation value $\langle \hat L\rangle$ and the minimal SEV $g_{\min}$ is attained.
	We find
	\begin{align}
		\langle\hat L\rangle=0.275
		\text{ and }
		g_{\mathrm{min}}=0.292,
	\end{align}
	which implies a successful entanglement test, $\langle \hat L\rangle<g_{\min}$ or $\langle \hat W\rangle=\langle \hat L\rangle-g_{\min}<0$.
	Note that the previously considered approaches \cite{Simon2000,Duan2000} did not verify this non-Gaussian form of entanglement.	
	The determined coherent amplitudes of our test operator in Eq. \eqref{eq:testoperator} can be given in the form
	\begin{align}\label{eq:a1}
		\alpha_1=Q_1,
		\text{}
		\alpha_2=1.2\,Q_2,
		\text{ and }
		\alpha_3=0.8\,Q_3,
	\end{align}
	as well as 
	\begin{align}\label{eq:b1}
		\beta_1=Q_1,
		\text{}
		\beta_2=0.8\,Q_2,
		\text{ and }
		\beta_3=1.2\,Q_3.
	\end{align}
	Here, the quantities $Q_k$ are defined as
	\begin{align}
	\begin{aligned}
		Q_1&=-\sqrt{2}\,\gamma
		\text{ and }\\
		Q_2&=\left[\left(\Delta+\dfrac1{\Delta}\right)+i\,\sqrt{\Delta^2+\dfrac1{\Delta^2}}\right]\dfrac{\gamma}{2}=Q_3^\ast,
	\end{aligned}
	\end{align}
	with $\Delta=\left(\sqrt{2}-1\right)^{1/3}$.
	Figure \ref{fig:confSupCoh} illustrates and summarizes the configurations of the different sets of complex amplitudes.
	
	The complex numbers $\{Q_1,Q_2,Q_3\}$ are pointed out here for the following reason.
	Let us choose the displacements to be $\alpha_k=\beta_k=Q_k$, instead of those in Eqs. \eqref{eq:a1} and \eqref{eq:b1}. 
	Then the SESs to the minimal SEV of the corresponding operator are the product states $|\gamma\rangle\otimes|-\gamma\rangle$ and $|-\gamma\rangle\otimes|\gamma\rangle$, which define the state under study; see also Appendix \ref{ch:specialprop}.

	As exemplified for the above case, we could have approached the verification of entanglement of state \eqref{eq:BellState} for all parameters $\gamma$ and $\epsilon$.
	Our non-Gaussian entanglement criteria in terms of displaced photon-number correlation are shown to outperform the applicability of the Simon and Duan \textit{et al.} approaches for the scenario under study.
	Furthermore, our optimization over the coherent amplitudes, defining our entanglement criteria, predicted their optimal choices for an experimental implementation of our technique.
	
\subsection{Two-mode squeezed-vacuum state}
\begin{figure}[ht]
	\includegraphics[clip,scale=0.55]{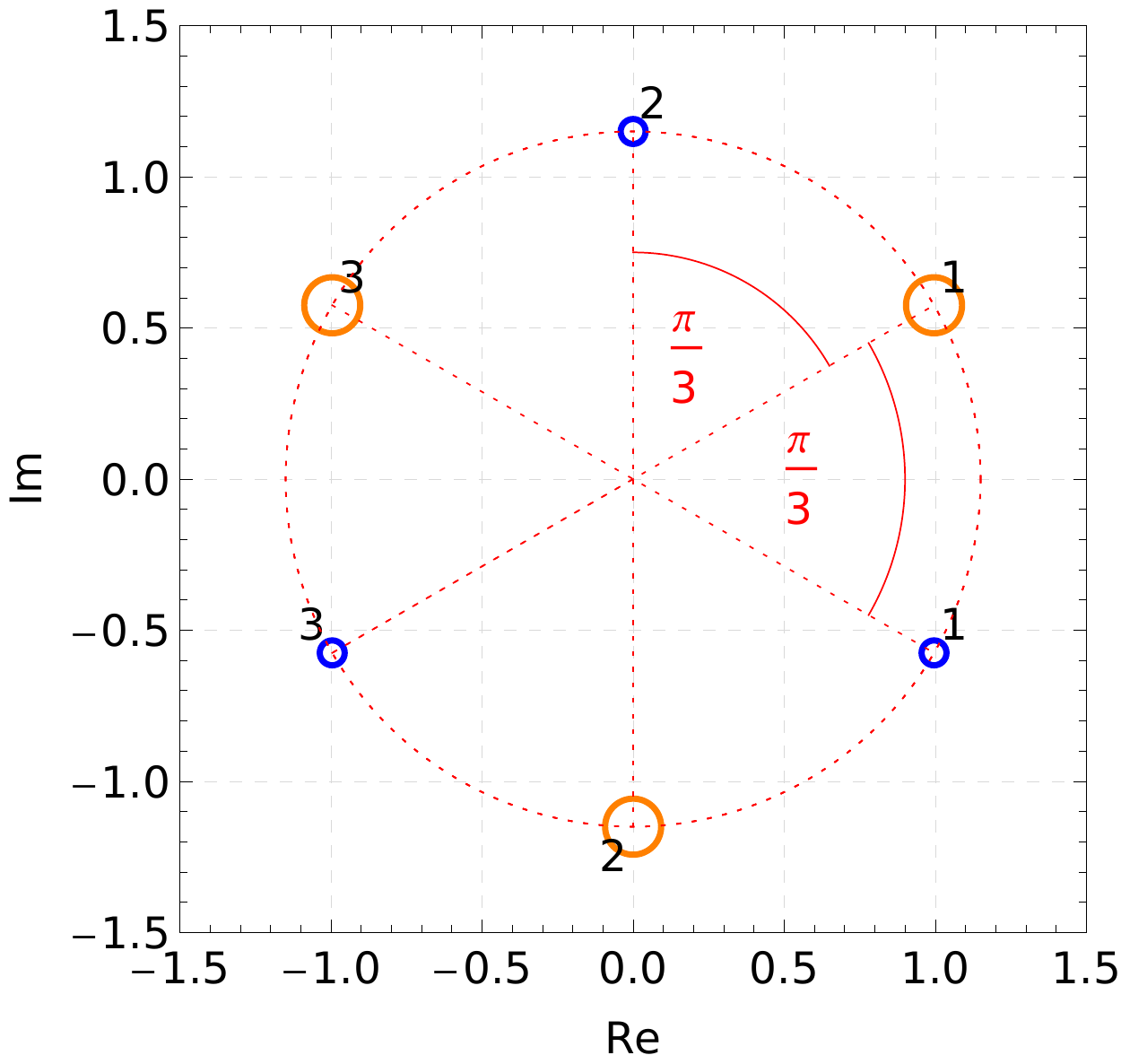}
	\caption{(Color online)
		Possible displacements $\alpha_k$ (larger, orange circles) and $\beta_k$ (smaller, blue circles) with $k=1,2,3$ for witnessing entanglement of the two-mode squeezed-vacuum state $|\xi\rangle$ are shown in the complex plane. Re and Im denote the real and imaginary axes, respectively.
	}\label{fig:confTMSV}
\end{figure}

	Let us now study the somewhat inverse scenario.
	That is, we apply our method to a Gaussian state.
	The underlying question is whether or not the measurement of correlated, displaced photon numbers allows one to detect Gaussian forms of entanglement.
	Thus, our second example is a two-mode squeezed-vacuum state,
	\begin{align}
		|\xi\rangle=\exp[-\xi\hat a^{\dagger}\otimes\hat b^\dagger+\xi^*\hat a\otimes\hat b]|0\rangle\otimes|0\rangle,
	\end{align}
	with complex squeezing parameter $\xi$.
	To apply our method, the expectation value of the displaced photon-number correlations is required,
	\begin{align}
        \begin{aligned}
		&\langle\xi|\hat n(\alpha_k)\otimes\hat n(\beta_k)|\xi\rangle\\
		=&\big(\sinh^2|\xi|+|\alpha_k|^2\big)\big(\sinh^2|\xi|+|\beta_k|^2\big)\\
		&+\left|\dfrac1{2}\sinh|2\xi|\,e^{i\arg(\xi)}-\alpha_k\beta_k\right|^2-|\alpha_k|^2|\beta_k|^2.
	\end{aligned}
	\end{align}
	For $\xi=0.5$, a parameter configuration is shown in Fig. \ref{fig:confTMSV}.
	In particular, we put the displacements $\{\alpha_1,\alpha_2,\alpha_3\}$ and $\{\beta_1,\beta_2,\beta_3\}$ to be on a circle of radius $r=1.15$, with $\alpha_k=re^{i[1/2-2(k-1)]\pi/3}$ and $\beta_k=\alpha_k^*$.
	The corresponding test operator $\hat L$ yields, for the given state, an expectation value which is significantly smaller than the minimal SEV,
	\begin{align}
		\langle\hat L\rangle=1.34
		\text{ and }
		g_{\mathrm{min}}=1.76.
	\end{align}
	Therefore, the Gaussian entanglement is verified.

\begin{figure}[ht]
	(a)\\
	\includegraphics[clip,scale=0.45]{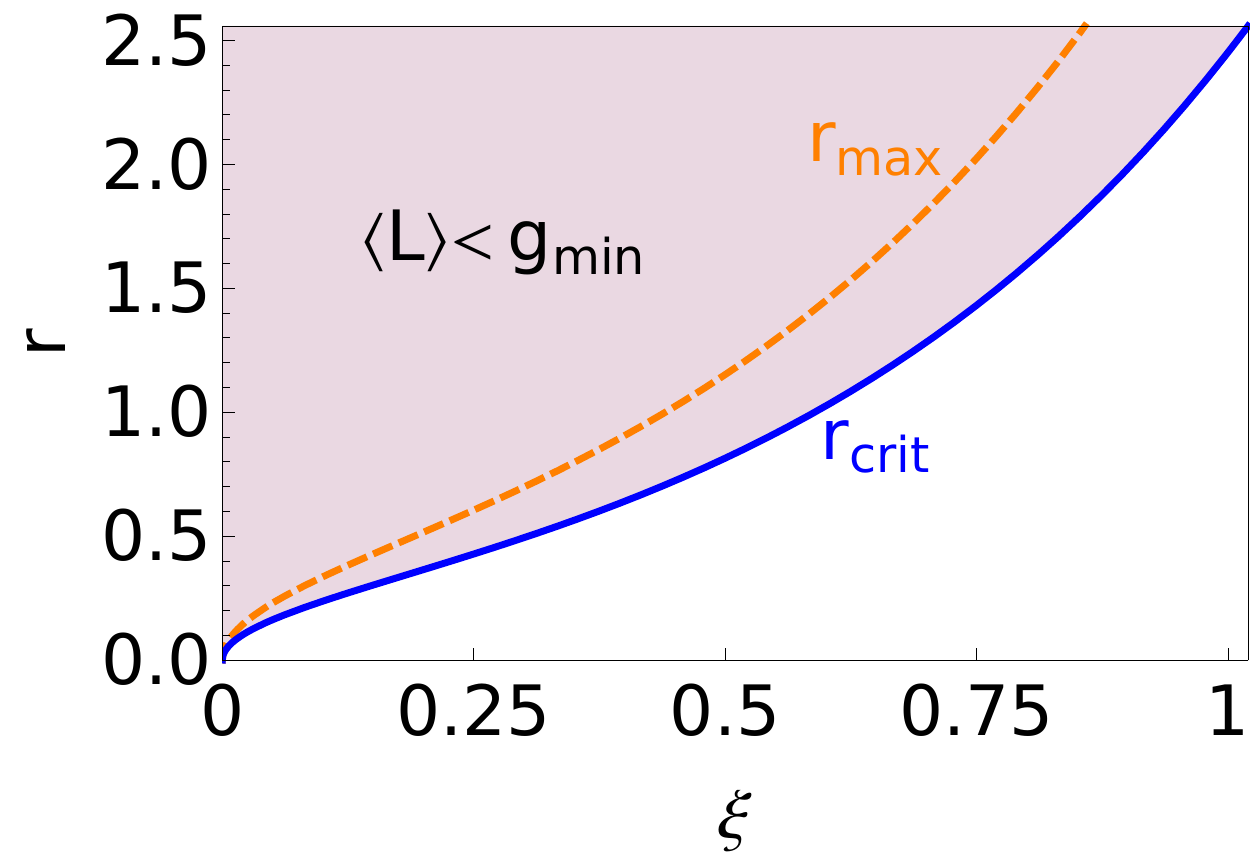}
	\\(b)\\
	\includegraphics[clip,scale=0.45]{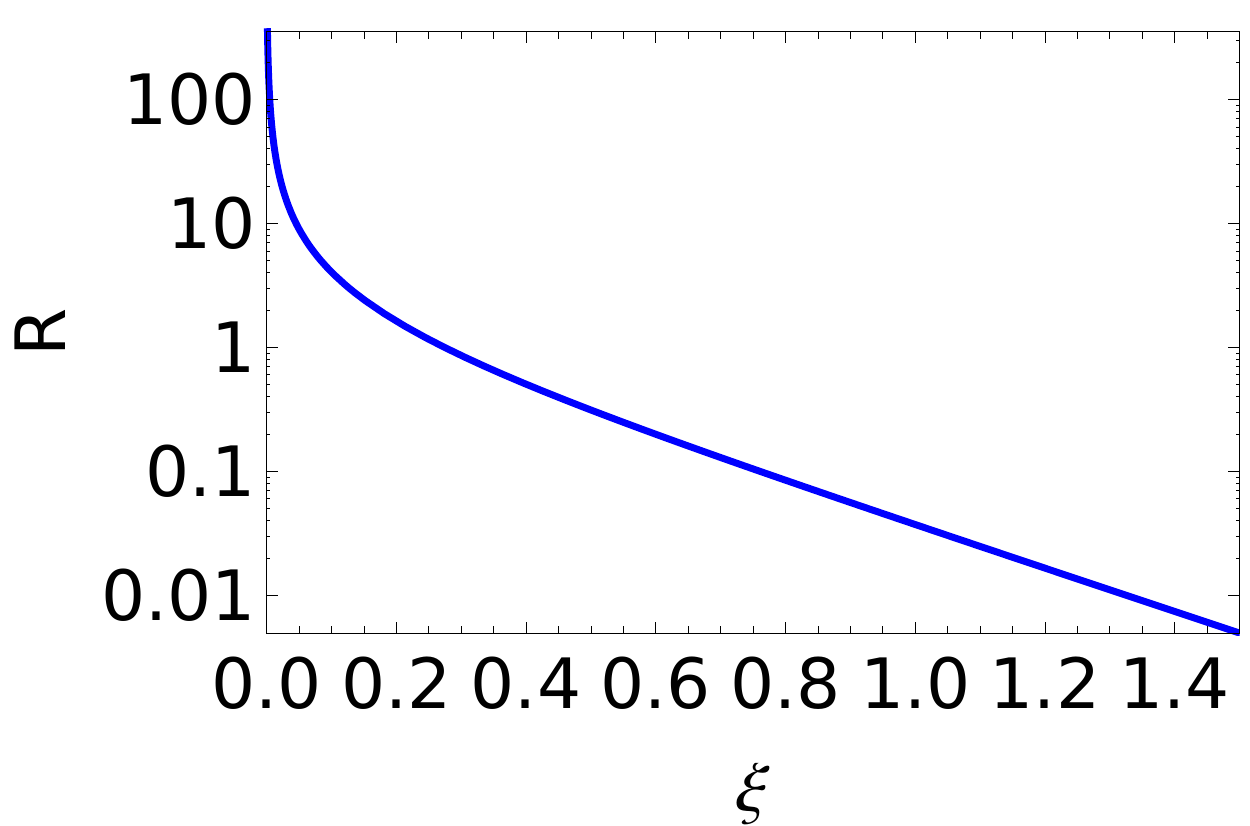}
	\caption{(Color online) 
		Top panel (a): The critical radius $r_\mathrm{crit}$ (solid, blue line) [Eq. \eqref{eq:CriticalRadius}] as a function of the amount of squeezing $\xi$.
		The shaded area corresponds to possible radii, which successfully certify entanglement.
 		The dashed, orange line shows the radius $r_{\max}$ [Eq. \eqref{eq:OptimalRadius}] for which $R$ [Eq. \eqref{eq:Relative}] is maximal.
		Bottom panel (b): $R$ on a logarithmic scale for $r_{\max}(\xi)$.
	}\label{fig:rc}
\end{figure}

	In order to get a better understanding of the relation between the test operator---more precisely, the configuration of coherent amplitudes---and the amount of squeezing, we vary the radius $r$.
	Note that the phases of the coherent amplitudes $\alpha_k$ and $\beta_k$ can be simply adjusted to $\arg(\xi)$ via corresponding rotations.
	It turns out that entanglement can be uncovered for any radius $r$ larger than the lower bound
	\begin{align}\label{eq:CriticalRadius}
		r_\mathrm{crit}(\xi)&=\dfrac1{2}\sqrt{\cosh(2|\xi|)\left(e^{2|\xi|}-1\right)},
	\end{align}
	which depends on the squeezing parameter.
	This relation is illustrated in Fig. \ref{fig:rc} (a) together with the radii,
	\begin{align}\label{eq:OptimalRadius}
	        r_{\max}(\xi)=\sqrt{2}\,r_\mathrm{crit}(\xi),
	\end{align}
	for which the positive-valued relative entanglement-detection quantity,
	\begin{align}\label{eq:Relative}
		R= \frac{g_{\mathrm{min}}}{\langle\hat L\rangle}-1,
	\end{align}
	is maximal.
	This maximal relative entanglement detection is shown as a function of $\xi$ in Fig. \ref{fig:rc} (b).
	Note that for the considered example $\xi=0.5$ the radius $r$ was chosen to be equal to $r_{\mathrm{max}}$.
	On the one hand, $R$ increases with decreasing $\xi$, which means that the relative resolution of detected entanglement is better for smaller squeezing levels.
	On the other hand, one observes for a weakly squeezed state that the absolute effect, 
	$g_{\mathrm{min}}-\langle\hat L\rangle$, becomes arbitrarily small.
	To relate this to the analysis of an experiment, let us mention that the absolute detection influences the measurement time or statistical significance, whereas the relative effect relates to the sensitivity or resolution of the employed measurement system.

\subsection{Single-photon-subtracted two-mode squeezed-vacuum state}
	As a final example for the bipartite scenario, we now consider another non-Gaussian state.
	In particular, we study a coherent single-photon-subtracted two-mode squeezed-vacuum state,
	\begin{align}
		|\psi_-\rangle\propto(\sqrt{\kappa}\,\hat a\otimes\hat{1}+\sqrt{1-\kappa}\,\hat{1}\otimes\hat b)|\xi\rangle,
	\end{align}
	with $\xi=0.5$.
	The parameter $0\leq\kappa\leq 1$ controls the relative amount of subtraction between the two modes.
	For further details on photon subtraction and its experimental implementation, see, e.g., Ref. \cite{Averchenko2016}.
	This example demonstrates how de-Gaussification processes may influence the detected entanglement.

	One readily derives for the displaced photon-number correlations the expression
	\begin{align}\nonumber
		&\langle\psi_-|\hat n(\alpha_k)\otimes\hat n(\beta_k)|\psi_-\rangle
		\\\nonumber
		=&6\sinh^4|\xi|+2\sinh^2|\xi|\,(|\alpha_k|^2+|\beta_k|^2+2)+|\alpha_k|^2|\beta_k|^2
		\\\nonumber
		&+\kappa|\alpha_k|^2+(1-\kappa)|\beta_k|^2-2\sinh|2\xi|e^{i\arg(\xi)}\mathrm{Re}(\alpha_k\beta_k)
		\\
		&+2\sqrt{\kappa(1-\kappa)}\mathrm{Re}(\alpha_k\beta_k^*)(1+2\sinh^2|\xi|). 
	\end{align}
	When the subtraction is performed in a balanced manner, $\kappa=1/2$, a suitable configuration of coherent amplitudes for the test operator, \eqref{eq:testoperator}, is given by $\alpha_1=re^{i\Theta}$, $\alpha_2=-ir$, $\alpha_3=-re^{-i\Theta}$, and $\beta_k=\alpha_k^\ast$, where $\Theta=\pi/5$ and $r=2.2$.
	For those amplitudes, we obtain
	\begin{align}
		\langle\hat L\rangle=22.72
		\text{ and }
		g_{\mathrm{min}}=22.98,
	\end{align}
	which verifies the entanglement, $\langle L\rangle<g_{\min}$, of the squeezed state subjected to a global (i.e., $0<\kappa<1$) photon-subtraction process.

	The configuration of coherent amplitudes is qualitatively different if the photon is removed locally.
	For example, a subtraction in the first mode, $\kappa=1$, yields the following coherent amplitudes for a successful entanglement test.
	They are $\alpha_1=r_{a}e^{i\pi/3}$, $\alpha_2=\alpha_1^\ast$, $\alpha_3=-r_{a}$, and $\beta_{k}=(r_{b}/r_{a})\alpha_{k}^\ast$, with $r_{a}=1.6$ and $r_{b}=2.2$.
	We get
	\begin{align}
		12.22=\langle\hat L\rangle<
		g_{\mathrm{min}}=12.39.
	\end{align}
	Note that due to symmetry, one can exchange $\alpha_{k}$ and $\beta_{k}$ in order to verify entanglement of the state where the photon is subtracted from the second mode, $\kappa=0$.

	Hence, the coherent amplitudes to be measured strongly depend on the prepared state.
	Using our technique, we can predict these values for efficient implementation and optimal entanglement detection.
	This example concludes our study of bipartite entanglement.

\section{Multimode entanglement detection}\label{ch:Multimode}
	In final section, we extend our analysis to multimode systems.
	Since many findings can be straightforwardly generalized from the bipartite case, we focus on the parts which differ from our previous considerations.
	Especially, we study entanglement for different mode partitions, i.e., instances of partial entanglement.

	Let us consider an $N$-mode system given in terms of the bosonic annihilation and creation operators $\hat a^{(j)}$ and $\hat a^{(j)\dag}$, respectively.
	The mode index $j$ is an element of the set $\mathcal{I}=\{1,\dots,N\}$.
	The displaced photon-number operator of the $j$th mode reads
	\begin{align}\label{eq:IndiDisplacedPN}
		\hat n^{(j)}(\alpha^{(j)})=\big(\hat a^{(j)}-\alpha^{(j)}\big)^\dag\big(\hat a^{(j)}-\alpha^{(j)}\big),
	\end{align}
	for a coherent amplitude $\alpha^{(j)}$.
	The complex amplitudes may be put into a vector, $\boldsymbol \alpha=(\alpha^{(1)},\ldots,\alpha^{(N)})^\mathrm{T}\in\mathbb C^{N}$.
	
	The entanglement properties of such a multimode system can be defined as follows.
	A $K$-partition is a decomposition of the set $\mathcal{I}$ into $K$ subsets $\mathcal{I}^{(\ell)}$ ($\ell=1,\dots,K$ and $1\leq K\leq N$).
	This means that the modes with an index in $\mathcal I^{(\ell)}$ are considered to be a joint subsystem.
	Now, a quantum state is separable with respect to the given partition, $\mathcal{I}^{(1)}:\cdots:\mathcal{I}^{(K)}$, if it can be written in the form
	\begin{align}
		\hat\sigma
		=\int dP(a^{(1)},\dots,a^{(K)})
		\bigotimes_{\ell=1}^K
		|a^{(\ell)}\rangle\langle a^{(\ell)}|,
	\end{align}
	where $|a^{(\ell)}\rangle\in\bigotimes_{j\in\mathcal{I}^{(\ell)}}\mathcal{H}_{j}$ and $P$ is a probability distribution.
	If such a representation is not possible, the state is entangled with respect to this partition.

	It is worth mentioning that a one-partition, $K=1$ or $\mathcal I^{(1)}=\mathcal I$, is referred to as a trivial partition, because there is no separation into different subsystems.
	A full partition, $K=N$ or $\mathcal I^{(\ell)}=\{\ell\}$ ($\ell=1,\ldots,N$), is the maximally possible decomposition.
	The intermediate levels of separation, $1<K<N$, result in a plethora of forms of partial separability.

\subsection{Probing displaced photon-number entanglement}
	We may first consider the measurement of displaced photon-number correlations.
	The displaced total photon number of the $\ell$th subsystem is given by the operator
	\begin{align}\label{eq:Nred}
		\hat N^{(\ell)}(\boldsymbol\alpha)=\sum_{j\in\mathcal I^{(\ell)}}q^{(j)}\hat n^{(j)}(\alpha^{(j)}),
	\end{align}
	where the weighting $q^{(j)}\geq0$ can be chosen according to some preferences to be specified.
	For instance, it may account for different detection efficiencies of the individual modes.
	Similarly to the bipartite scenario, we aim at detecting entanglement with an operator of the form
	\begin{align}\label{eq:discreteL}
		\hat L=\sum_{k=1}^m\lambda_k \hat N^{(1)}(\boldsymbol\alpha_k)\otimes\cdots\otimes\hat N^{(K)}(\boldsymbol\alpha_k),
	\end{align}
	for the given $K$-partition $\mathcal{I}^{(1)}:\cdots:\mathcal{I}^{(K)}$ and $m$ different multimode displacements $\boldsymbol\alpha_k\in\mathbb C^{N}$ for $k=1,\ldots,m$.
	Again, we restrict ourselves to positive coefficients $\lambda_k$ with $\sum_k\lambda_k=1$.
	The operator, \eqref{eq:discreteL}, is of the order $2K$ in terms of the creation and annihilation operators.
	Thus, except for the trivial partition $K=1$, the expectation value $\langle\hat L\rangle$ depends on non-Gaussian characteristics of the state under study.

\begin{figure}[hb]
	\includegraphics[clip,scale=0.55]{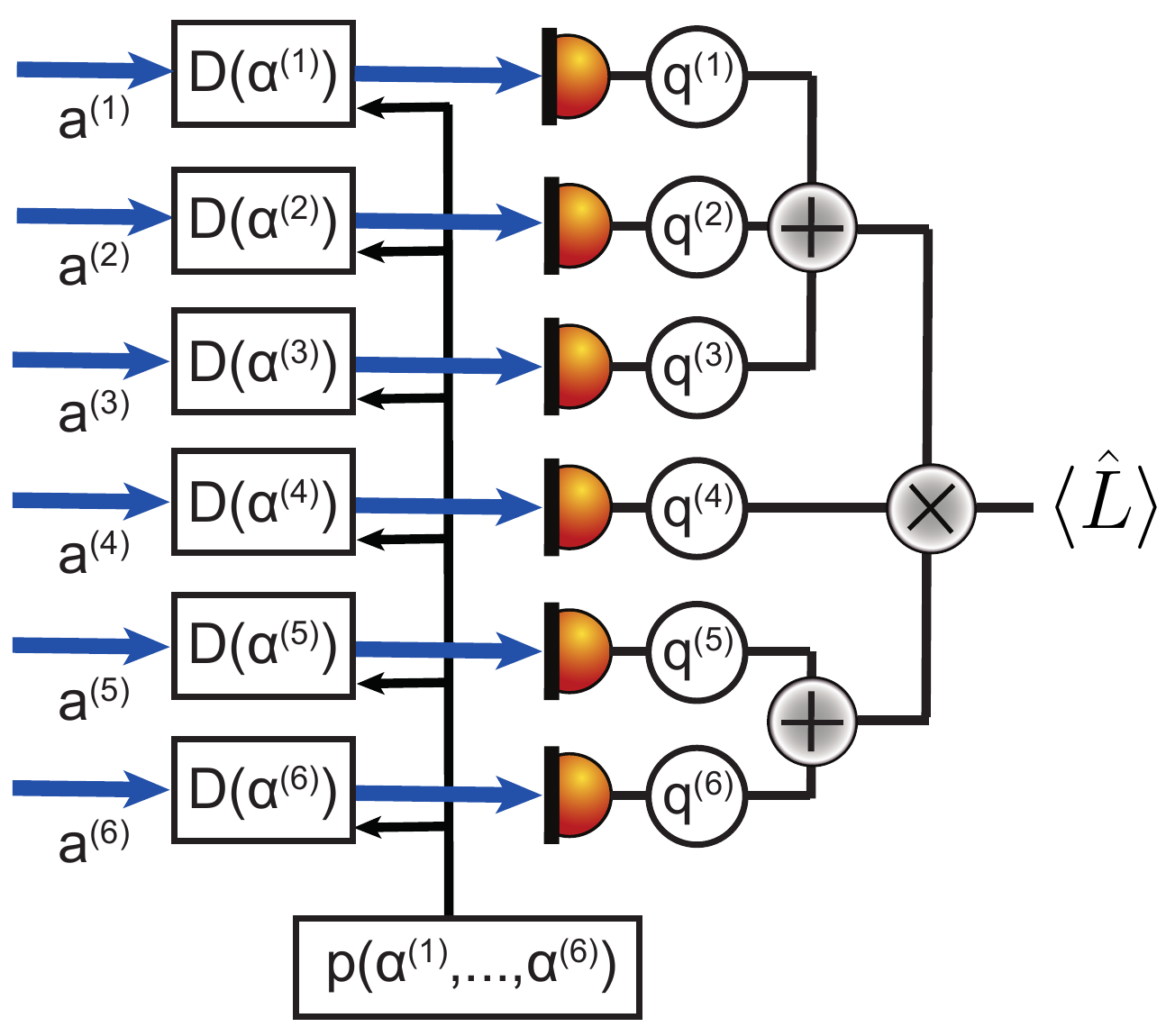}
	\caption{(Color online)
		Experimental setup to measure the observable, \eqref{eq:discreteL}, for the example of the tripartition $\{1,2,3\}:\{4\}:\{5,6\}$.
		Compare also to the bipartite case in Fig. \ref{fig:DNESetup}.
		The displacements of the individual modes are randomly realized.
		The measurement of the resulting displaced photon-number statistics is combined in a weighted sum for each subsystem---using the weights $q^{(1)},\ldots,q^{(6)}$---and is then correlated.
	}\label{fig:expmulti}
\end{figure}

	The experimental measurement strategy has to be adapted to a given partition in order to infer the observable $\hat L$ [Eq. \eqref{eq:discreteL}].
	This is illustrated in Fig. \ref{fig:expmulti} for the tripartition $\{1,2,3\}:\{4\}:\{5,6\}$ of a six-mode system.
	Each mode is coherently displaced, $\hat D(\boldsymbol{\alpha})=\bigotimes_{j=1}^N\hat D(\alpha^{(j)})$, where $\boldsymbol{\alpha}$ is one of the possible realizations $\boldsymbol{\alpha}_k$, $k=1,\dots,m$.
	The photon number of the transmitted beam is recorded afterwards.
	For each subsystem of the partition, $\{1,2,3\}$, $\{4\}$, and $\{5,6\}$, the total photon number [Eq. \eqref{eq:Nred}] is determined by summing the detector outcomes of all modes of the subset using the weights $q^{(j)}$.
	Subsequently, the results for the individual subsystems are multiplied.
	Random displacements of the modes by amplitudes $\boldsymbol\alpha_k$ are performed with probabilities $\lambda_k$.
	For sufficiently long data acquisition times, the measurement outcome approaches the expectation value of the operator, \eqref{eq:discreteL}.

\subsection{Multimode separability eigenvalue problem}
	In close analogy to the previously studied bipartite scenario, inseparability with respect to a given partition can be probed via the conditions \cite{Sperling2013}
	\begin{align}\label{eq:MultiEntCond}
		\langle \hat L\rangle<g_{\min}^{\mathcal{I}^{(1)}:\cdots:\mathcal{I}^{(K)}}.
	\end{align}
	Here, $g_{\min}^{\mathcal{I}^{(1)}:\cdots:\mathcal{I}^{(K)}}$ is the minimal expectation value of $\hat L$ for states which are separable with respect to the considered $K$ partition.
	Again, this bound is the minimal SEV of the multipartite generalization of the SEEs \eqref{eq:SEPintroa} and \eqref{eq:SEPintrob}.
	The multimode SEEs read \cite{Sperling2013}
	\begin{align}\label{eq:multimodeSEE}
		\hat L_{a^{(1)},\dots,a^{(j-1)},a^{(j+1)},\dots,a^{(K)}}|a^{(j)}\rangle=g|a^{(j)}\rangle,
	\end{align}
	for $j=1,\dots,K$.
	The operator on the left-hand side of Eq. \eqref{eq:multimodeSEE} is the reduced operator with respect to all but the $j$th subsystem; see Eqs. \eqref{eq:ReducedOpA} and \eqref{eq:ReducedOpB} for the bipartite case.

	For example, the reduced operators for the specific observable in Eq. \eqref{eq:discreteL} are
	\begin{align}\label{eq:ParLred}
	\begin{aligned}
		&\hat L_{a^{(1)},\dots,a^{(j-1)},a^{(j+1)},\dots,a^{(K)}}\\
		=&\sum_{k=1}^m\lambda_k
		\prod_{\substack{\ell=1\\ \ell\neq j}}^K\langle a^{(\ell)}|\hat N^{(\ell)}_k(\boldsymbol\alpha_k)|a^{(\ell)}\rangle
		\hat N^{(j)}_k(\boldsymbol\alpha_k),
	\end{aligned}
	\end{align}
	for $1\leq j\leq K$.
	For all subsystems, they have the general form of a sum of displaced photon-number operators,
	\begin{align}\label{eq:multimodeLred}
		\hat L'=\sum_{k} \Lambda_k\sum_{j}q^{(j)} \hat n^{(j)}(\alpha^{(j)}_k),
	\end{align}
	with nonnegative coefficients $\Lambda_k$ and complex displacements $\alpha_{k}^{(j)}$.
	This, analogously to the bipartite case, implies that the multipartite SES to the minimal SEV is a multimode coherent state, $|\beta^{(1)}\rangle\otimes\cdots\otimes|\beta^{(N)}\rangle$.
	In fact, it is worth mentioning that the number of superpositions $m$ in Eq. \eqref{eq:discreteL} should exceed the number of partitions $K$ by at least one, $m\geq K+1$, to have a useful entanglement witness.

	Moreover, this also implies for the multipartite SEEs, \eqref{eq:multimodeSEE}, that the coherent amplitudes have to obey certain relations.
	Those can be analogously formulated as done in the bipartite case in Sec. \ref{subsec:DPNC2mode}.
	Furthermore, the minimal multipartite SEV can be obtained numerically as it was discussed in Sec. \ref{subsec:SpecialCases}.
	In addition, and for simplicity, we use the factors $q^{(j)}=1/|\mathcal{I}^{(\ell)}|$ ($j\in\mathcal I^{(\ell)}$ and $|\mathcal X|$ is the cardinality of the set $\mathcal X$) for the following examples.

\subsection{Example: Multimode mixed states}\label{ch:multimodeexamples}
	To demonstrate the general capabilities of our technique, we may begin with a non-Gaussian four-mode ($N=4$) state,
	\begin{align}\label{eq:fourgamma}
		|\psi_\gamma\rangle=\frac{
			|\gamma,\gamma,\gamma,\gamma\rangle+|-\gamma,-\gamma,-\gamma,-\gamma\rangle
		}{
			\sqrt{2\left(1+e^{-8|\gamma|^2}\right)}
		},
	\end{align}
	which is a quantum superposition of two coherent states.
	Such a state can be produced experimentally by splitting a single-mode cat state, being a quantum superposition of the coherent states $|2\gamma\rangle$ and $|-2\gamma\rangle$, on a 4-splitter.
	For the generation of the cat state, we refer, for example, to Ref. \cite{Jeong2014}.

	In addition, we assume that the coherent amplitude $\gamma$ is not perfectly determined.
	Assuming Gaussian noise,
	\begin{align}\label{eq:GaussNoise}
		P_\gamma(\gamma')=\dfrac1{2\pi\sigma^2}\exp\left[-\dfrac{|\gamma'-\gamma|^2}{2\sigma^2}\right],
	\end{align}
	we get the mixed non-Gaussian four-mode state
	\begin{align}\label{eq:MultiMixedState}
		\hat\rho_{\gamma,\sigma}=\int d^2\gamma'\,P_\gamma(\gamma')|\psi_{\gamma'}\rangle\langle \psi_{\gamma'}|.
	\end{align}
	We are going to study the entanglement of this state.
	Due to symmetry, it is sufficient to consider the four-partition $\{1\}:\{2\}:\{3\}:\{4\}$, the tripartition $\{1\}:\{2,3\}:\{4\}$, and the bipartitions $\{1\}:\{2,3,4\}$ and $\{1,2\}:\{3,4\}$ only.
	Also, we particularly investigate the case $\gamma=0.4$.

	Again, we apply the genetic optimization algorithm to find the optimal parameters for entanglement certification by means of the test operator defined in Eq. \eqref{eq:discreteL} together with Eq. \eqref{eq:Nred}.
	This shows that one can restrict to coherent displacements, $\alpha^{(j)}_k$ for $j=1,\dots,N$ and $k=1,\dots,m$, on the line in phase space that connects $\gamma$ with $-\gamma$.
	This can greatly simplify the experimental implementation.
	In fact, we provide a suitable parameter configuration for all possible partitions of the state under study in Appendix \ref{ch:examplemultimode} together with the minimal SEV and the expectation value of the test operator for state \eqref{eq:fourgamma}.

\begin{figure}[ht]
	\centering
	\begin{minipage}[t]{0.49\linewidth}
		(a)\\
		\includegraphics[width=0.95\linewidth]{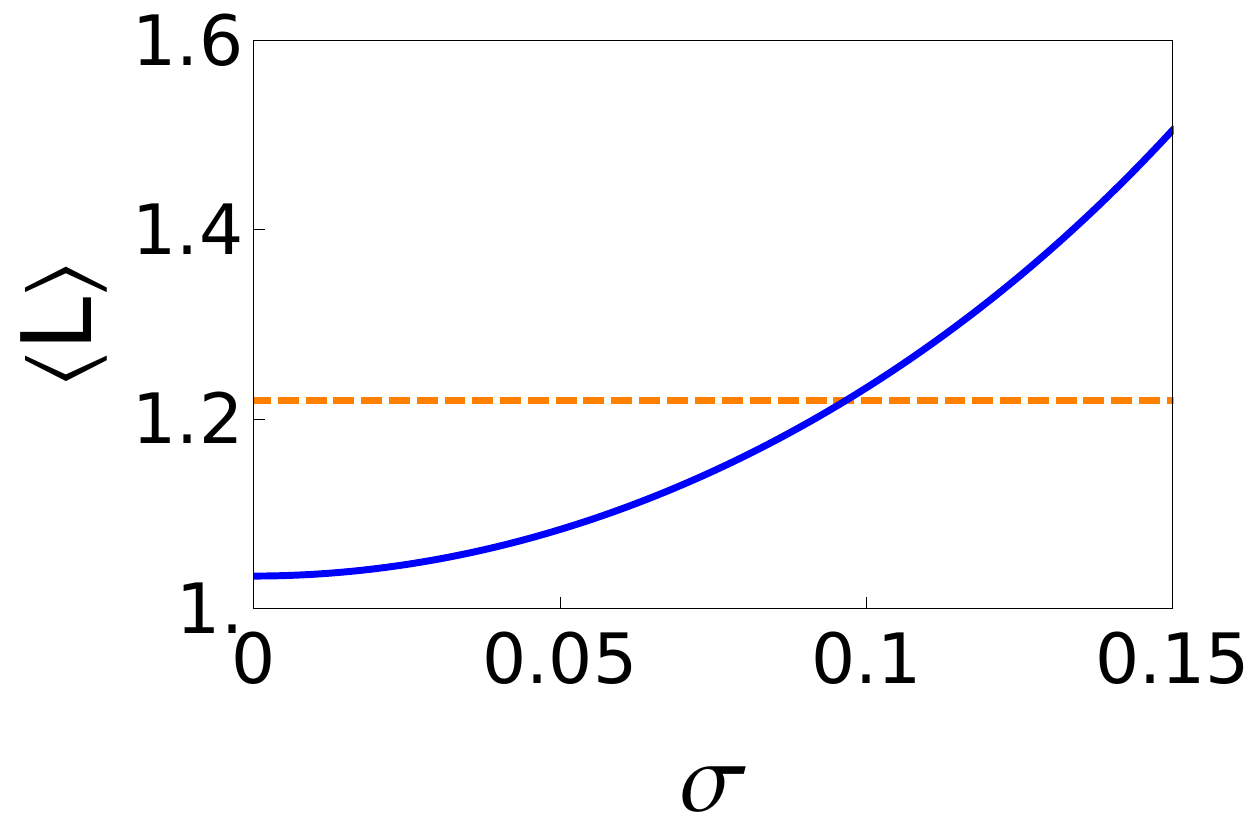}
		\label{fig:noisycat1234}
		\\(c)\\
		\includegraphics[width=0.95\linewidth]{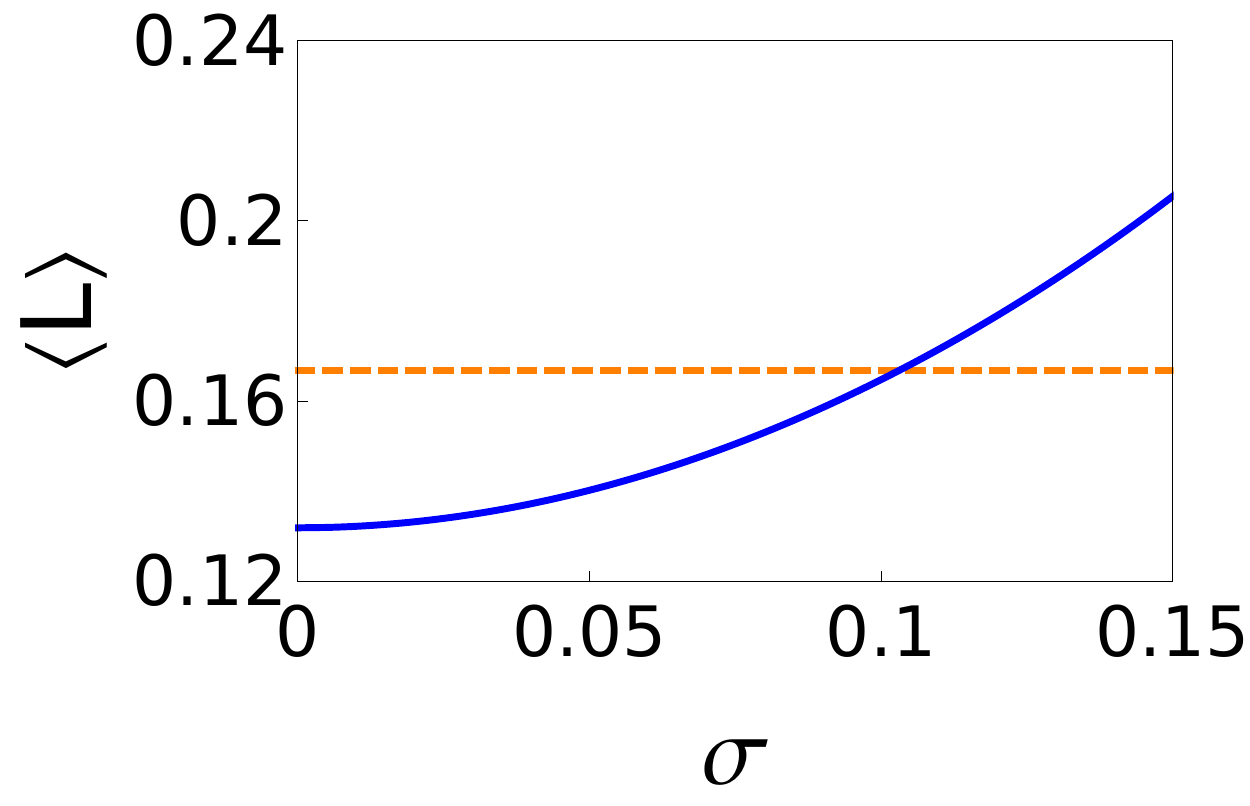}
		\label{fig:noisycat1122}
	\end{minipage}
	\hfill 
	\begin{minipage}[t]{0.49\linewidth}
		(b)\\
		\includegraphics[width=0.95\linewidth]{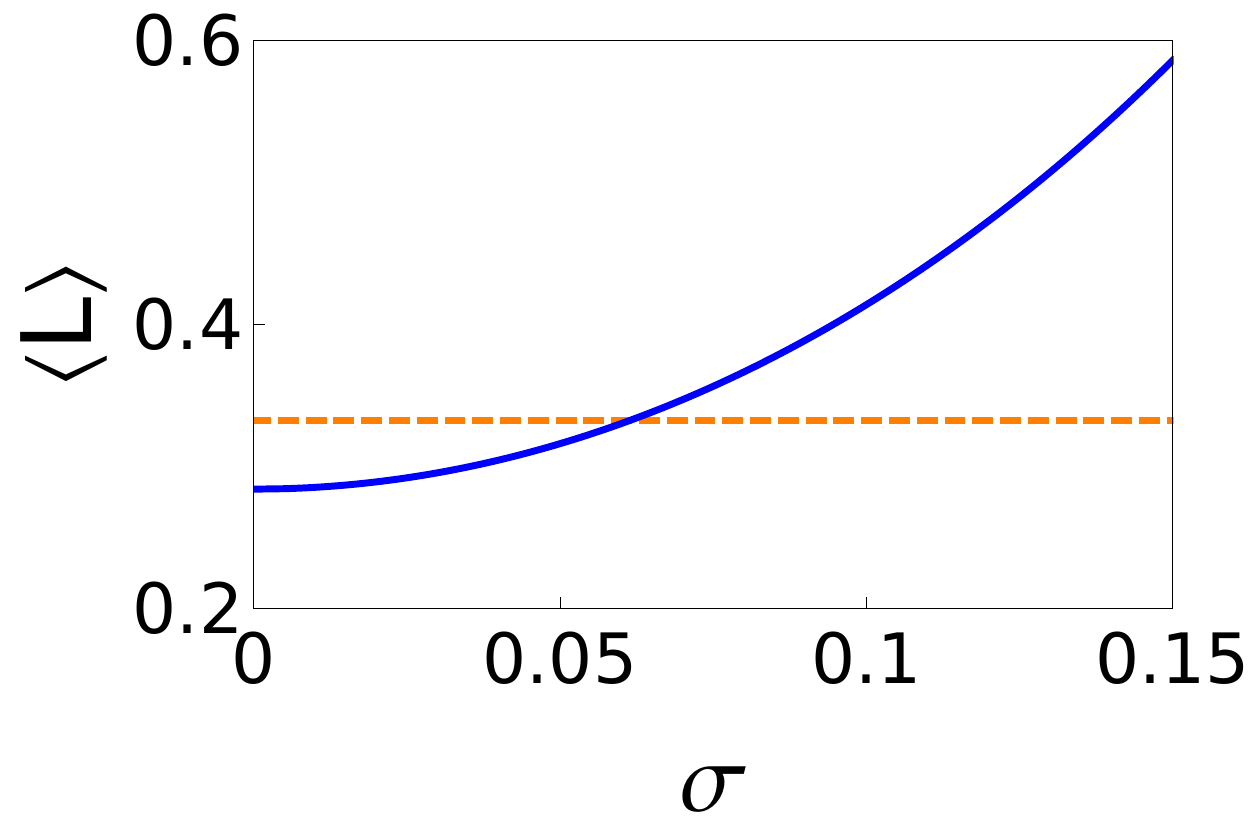}
		\label{fig:noisycat1223}
		\\(d)\\
		\includegraphics[width=0.95\linewidth]{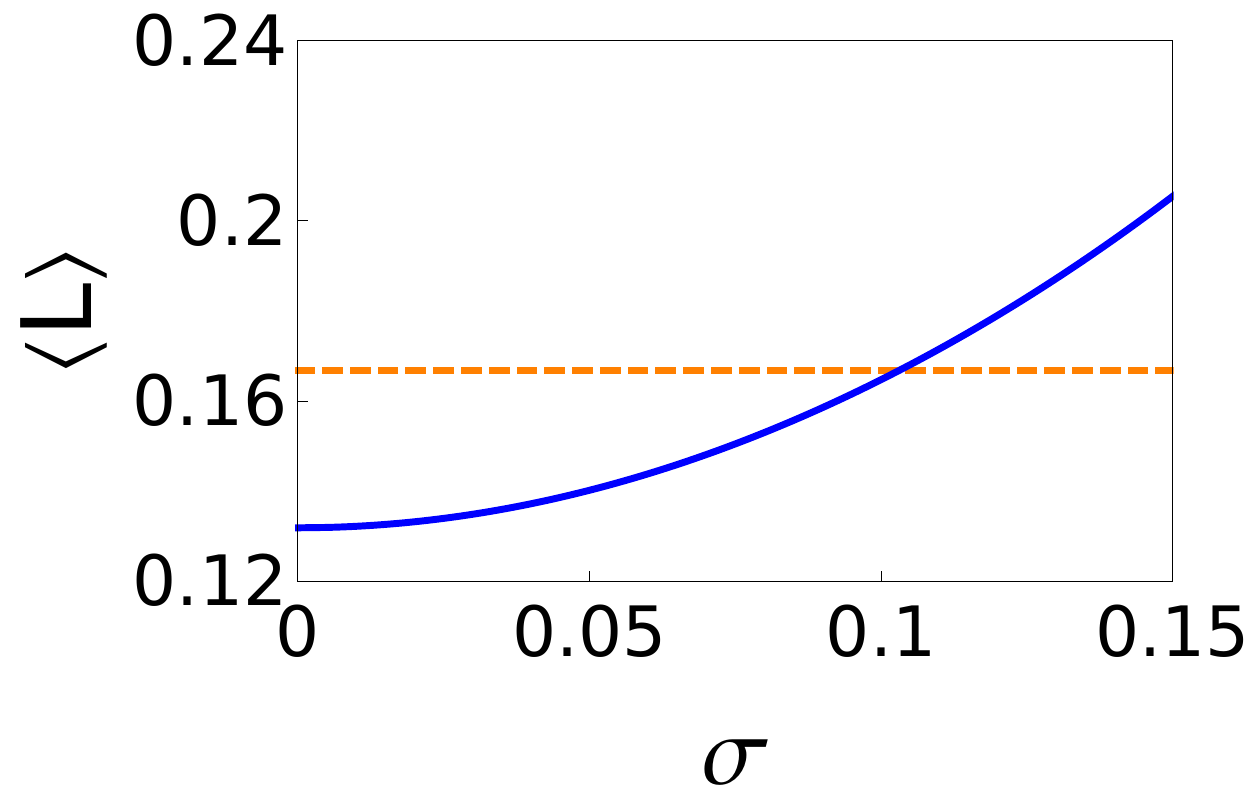}
		\label{fig:noisycat1222}
	\end{minipage}
	\caption{
		(Color online)
		Expectation value $\langle L\rangle$ for different $K$ partitions, $\mathcal I^{(1)}:\cdots:\mathcal I^{(K)}$ as a function of the noise  parameter $\sigma$ (solid blue curves).
		We study (a) $\{1\}:\{2\}:\{3\}:\{4\}$;
		(b) $\{1\}:\{2,3\}:\{4\}$;
		(c) $\{1,2\}:\{3,4\}$;
		(d) $\{1\}:\{2,3,4\}$.
		The bounds for $K$ separability, $g_{\min}^{\mathcal I^{(1)}:\cdots:\mathcal I^{(K)}}$, are represented as dashed orange lines.
		Entanglement is verified for $\langle L\rangle<g_{\min}^{\mathcal I^{(1)}:\cdots:\mathcal I^{(K)}}$; see entanglement condition \eqref{eq:MultiEntCond}.
	}
	\label{fig:noise}
\end{figure}

	Figures \ref{fig:noise}(a)--\ref{fig:noise}(d) show the expectation value $\langle\hat L\rangle$ for state \eqref{eq:MultiMixedState} ($\gamma=0.4$) as a function of the standard deviation $\sigma$ of the Gaussian noise, \eqref{eq:GaussNoise}, together with the lower bound of expectation values for separable states.
	For noise levels below a critical standard deviation, $\sigma<\sigma_\mathrm{crit}$, entanglement is uncovered.
	The critical standard deviations for the individual partitions are listed in Table \ref{tab:sigc}.
	Thus, our entanglement verification approach not only is able to detect different instances of multimode entanglement, but also is robust---to the extent discussed---against the considered forms of imperfections.

\begin{table}[ht]
	\centering
	\caption{
		Critical standard deviations $\sigma_\mathrm{crit}$ of the Gaussian noise for the individual $K$ partitions.
		Entanglement is verified for noise levels below those values.
	}\label{tab:sigc}
	\begin{tabular}{c c c}
		\hline\hline
		$K$ & \hspace*{1cm} Partition \hspace*{1cm} & $\sigma_\mathrm{crit}$
		\\\hline
		4&$\{1\}:\{2\}:\{3\}:\{4\}$ & 0.097
		\\
		3&$\{1\}:\{2,3\}:\{4\}$ & 0.061
		\\
		2&$\{1,2\}:\{3,4\}$ & 0.103
		\\
		2&$\{1\}:\{2,3,4\}$ & 0.103
		\\\hline\hline
	\end{tabular}
\end{table}

\section{Conclusions}\label{ch:Conclusions}
	Motivated by recent measurement schemes capable of inferring photon numbers in a phase-sensitive manner, we construct entanglement criteria which are based on displaced photon-number correlations of multimode radiation fields.
	Our family of entanglement conditions is formulated in terms of observables which are combinations of photon-number correlations for different displacements.
	Using the method of separability eigenvalue equations, we determine the lower bounds of the expectation values of these observables for separable states, whose violation infers entanglement.

	We apply our approach to study entanglement of bipartite systems as well as different instances of multipartite entanglement.
	Applying a genetic optimization algorithm, we find the observable which yields an optimal entanglement detection for the state under study within the constructed family of observables.
	For example, we demonstrate for some considered non-Gaussian states that our entanglement tests are more sensitive than criteria which are typically employed.
	Also, different forms of partial entanglement of multipartite systems have been verified for the example of four-mode states.
	We are able to predict bounds to Gaussian noise for which entanglement remains detectable for this example.

	Furthermore, we compare our approach to verify entanglement with another concept of quantum correlation typically applied in quantum optics. 
	In addition, we include detection losses in our analysis.
	It is demonstrated that entanglement is detectable---independently of the amount of constant loss at each detector.

	Let us also mention that our technique is also applicable to an ensemble of trapped ions, whose vibrational motions are coupled \cite{Solano2002}, because this system has a structure mathematically similar to that of quantized radiation fields.
	The motional energy eigenstates of the ions can be individually probed by lasers via quantum nondemolition measurements \cite{deMatosFilho1996}.
	Coherent displacement of the motional states can be performed as well \cite{Ziesel2013}.
	Consequently, our method can be straightforwardly extended to this scenario of trapped ions.

	Thus, we present a versatile method to probe entanglement based on displaced photon-number correlations.
	Our analysis and the examples emphasize the strength and robustness of the method, and a generalization to trapped-ion systems has been outlined.
	Hence, we believe that our experimentally accessible technique will be helpful to further improve the understanding and the verification of the important class of non-Gaussian entanglement in complex systems, which is of great relevance for various applications in quantum technology.


\acknowledgments
	The authors are grateful to Stefan Gerke for enlightening discussions.
	This work was supported by the European Commission through the project QCUMbER (Quantum Controlled Ultrafast Multimode Entanglement and Measurement), Grant No. 665148.

\appendix

\section{Solving the separability eigenvalue problem for constant loss}\label{ch:appconstantloss}

	Here we consider imperfect detection, in particular, constant detection loss. 
	Let us start with the observable
	\begin{align}
		\hat L=\sum_{k=1}^m\lambda_k\hat n(\alpha_k)\otimes\hat n(\beta_k), 
	\end{align}
	which is measured with ideal detectors.
	Its minimal SEV is 
	\begin{align}\label{eq:gmin}
		g_{\min}=\sum_{k=1}^m\lambda_k|\alpha-\alpha_k|^2|\beta-\beta_k|^2,
	\end{align}
	and the corresponding SES is $|\alpha\rangle\otimes|\beta\rangle$, which is a two-mode coherent state.
	Its amplitudes obey the equation system
	\begin{subequations}\label{eq:nleq}
	\begin{align}
		\sum_{k=1}^m\lambda_k|\beta-\beta_k|^2(\alpha-\alpha_k)&=0,\\
		\sum_{k=1}^m\lambda_k|\alpha-\alpha_k|^2(\beta-\beta_k)&=0.		
	\end{align}
	\end{subequations}
	If mode $a$ and $b$ suffer constant loss, described by efficiencies $\eta_a$ and $\eta_b$, one measures the transformed operator
	\begin{align}
		\hat L^{(\eta_a,\eta_b)}=\eta_a\eta_b\sum_{k=1}^m\lambda_k\hat n\left(\dfrac{\alpha_k}{\sqrt{\eta_a}}\right)\otimes\hat n\left(\dfrac{\beta_k}{\sqrt{\eta_b}}\right). 
	\end{align}
	The SES corresponding to the minimal SEV of this operator is also a two-mode coherent state, $|\alpha^{(\eta_a,\eta_b)}\rangle\otimes|\beta^{(\eta_a,\eta_b)}\rangle$, with complex amplitudes $\alpha^{(\eta_a,\eta_b)}$ and $\beta^{(\eta_a,\eta_b)}$.
	Accordingly, the minimal SEV of $\hat L^{(\eta_a,\eta_b)}$ is given by
	\begin{align}\label{eq:gmineta}
		g^{(\eta_a,\eta_b)}_{\mathrm{min}}=\eta_a\eta_b\sum_{k=1}^m\lambda_k\left|\alpha^{(\eta_a,\eta_b)}-\dfrac{\alpha_k}{\sqrt{\eta_a}}\right|^2\left|\beta^{(\eta_a,\eta_b)}-\dfrac{\beta_k}{\sqrt{\eta_b}}\right|^2
	\end{align}
	and the unknown complex amplitudes $\alpha^{(\eta_a,\eta_b)}$ and $\beta^{(\eta_a,\eta_b)}$ follow from the coupled equation system
	\begin{subequations}
	\begin{align}
		\eta_a\eta_b\sum_{k=1}^m\!\lambda_k\!\left|\beta^{(\eta_a,\eta_b)}-\dfrac{\beta_k}{\sqrt{\eta_b}}\right|^2\!\!\left(\alpha^{(\eta_a,\eta_b)}-\dfrac{\alpha_k}{\sqrt{\eta_a}}\right)&\!=\!0,\\
		\eta_a\eta_b\sum_{k=1}^m\!\lambda_k\!\left|\alpha^{(\eta_a,\eta_b)}-\dfrac{\alpha_k}{\sqrt{\eta_a}}\right|^2\!\!\left(\beta^{(\eta_a,\eta_b)}-\dfrac{\beta_k}{\sqrt{\eta_b}}\right)&\!=\!0.
	\end{align}
	\end{subequations}
	It can be rewritten as
	\begin{subequations}
	\begin{align}
		\sum_{k=1}^m\!\lambda_k\!\left|\sqrt{\eta_b}\beta^{(\eta_a,\eta_b)}-\beta_k\right|^2\!\!\left(\sqrt{\eta_a}\alpha^{(\eta_a,\eta_b)}-\alpha_k\right)&\!=\!0,\\
		\sum_{k=1}^m\!\lambda_k\!\left|\sqrt{\eta_a}\alpha^{(\eta_a,\eta_b)}-\alpha_k\right|^2\!\!\left(\sqrt{\eta_b}\beta^{(\eta_a,\eta_b)}-\beta_k\right)&\!=\!0.
	\end{align}
	\end{subequations}
	Together with Eq. \eqref{eq:nleq}, it follows directly that
	\begin{subequations}
	\begin{align}
		\alpha^{(\eta_a,\eta_b)}&=\dfrac{\alpha}{\sqrt{\eta_a}},\\
		\beta^{(\eta_a,\eta_b)}&=\dfrac{\beta}{\sqrt{\eta_b}}.
	\end{align}
	\end{subequations}
	Inserting these amplitudes into Eq. \eqref{eq:gmineta}, one gets for the minimal SEV
	\begin{align}
		g^{(\eta_a,\eta_b)}_{\mathrm{min}}=\sum_{k=1}^m\lambda_k|\alpha-\alpha_k|^2|\beta-\beta_k|^2, 
	\end{align}
	which is obviously the same as the minimal SEV, $g_{\mathrm{min}}$, for the original operator $\hat L$ [cf. Eq. \eqref{eq:gmin}], i.e.,
	\begin{align}
	        g^{(\eta_a,\eta_b)}_{\mathrm{min}}=g_{\mathrm{min}}.
	\end{align}
	These considerations can be easily generalized to the multimode case.

\section{Special property}\label{ch:specialprop}
	Let us formulate a relation between a given state, the test operator $\hat L$, and its SESs to the minimal SEV.
	For any coherent amplitude $\gamma\in\mathbb{C}$ and arbitrary coherent displacements $\alpha_k\in\mathbb{C}$, one can easily verify that
	\begin{align}\label{eq:twosessup}
	\begin{aligned}
		&\langle\gamma,-\gamma|\hat n(\alpha_k)\otimes\hat n(\alpha_k)|-\gamma,\gamma\rangle\\
		=&\langle-\gamma,\gamma|\hat n(\alpha_k)\otimes\hat n(\alpha_k)|-\gamma,\gamma\rangle\langle\gamma,-\gamma|-\gamma,\gamma\rangle.
	\end{aligned}
	\end{align}
	Obviously, the following symmetry relation also holds true,
	\begin{align}\label{eq:twoses}
	\begin{aligned}
	       &\langle-\gamma,\gamma|\hat n(\alpha_k)\otimes\hat n(\alpha_k)|-\gamma,\gamma\rangle\\
	       =&\langle\gamma,-\gamma|\hat n(\alpha_k)\otimes\hat n(\alpha_k)|\gamma,-\gamma\rangle.
	\end{aligned}
	\end{align}
	Thus, we get for the state in Eq. \eqref{eq:BellState} that
	\begin{align}
	\begin{aligned}
		&\langle\psi|\hat n(\alpha_k)\otimes\hat n(\alpha_k)|\psi\rangle\\
		=&\langle-\gamma,\gamma|\hat n(\alpha_k)\otimes\hat n(\alpha_k)|-\gamma,\gamma\rangle.
	\end{aligned}
	\end{align}
	Now, we consider the test operator
	\begin{align}
		\hat L=\sum_{k=1}^m\lambda_k\hat n(\alpha_k)\otimes\hat n(\alpha_k) 
	\end{align}
	with the minimal separability eigenvalue $g_{\mathrm{min}}$.
	If the corresponding SES to $g_{\mathrm{min}}$ is $|\gamma,-\gamma\rangle$, which is the case for proper choice of the $\alpha_k$, then also $|-\gamma,\gamma\rangle$ is an SES to $g_{\mathrm{min}}$ [cf. Eq. \eqref{eq:twoses}].
	Moreover, due to Eq. \eqref{eq:twosessup}, each quantum superposition of $|\gamma,-\gamma\rangle$ and $|-\gamma,\gamma\rangle$ has the same expectation value $\langle\hat L\rangle=g_{\mathrm{min}}$ as $|\gamma,-\gamma\rangle$.

\section{Witness configurations}\label{ch:examplemultimode}
	Here we summarize, for all the partitions of the four-mode case studied in Sec. \ref{ch:multimodeexamples}, the properties of the determined test operators which allow us to certify entanglement.
	In particular, these are the coherent displacement amplitudes, $\{\alpha^{(j)}_k\}^{j=1,\dots,N}_{k=1,\dots,m}$, and weighting factors, $\{\lambda_k\}_{k=1,\dots,m}$, of the test operator in Eq. \eqref{eq:discreteL} together with Eq. \eqref{eq:Nred}. 
	Furthermore, the expectation value $\langle\hat L\rangle$ for state \eqref{eq:fourgamma} and the minimal separability eigenvalue $g_{\mathrm{min}}$ are specified.
	Note that we use the factors $q^{(j)}=1/|\mathcal{I}^{(\ell)}|$ ($j\in\mathcal I^{(\ell)}$).

	The columns of the matrices given below address the displacement configurations $k=1,\dots,m$, while the rows label the respective modes $j=1,\dots,N$.
	For the four-partition $\{1\}:\{2\}:\{3\}:\{4\}$, we have
	\begin{align}
	\begin{aligned}
		\{\alpha^{(j)}_k\}&=
		\begin{pmatrix}
		  -1.3&-0.3& 0.7& 1.7&2.7\\
		  -2.3&-1.3&-0.3& 0.7& 1.7\\
		  0.3& 1.3&-2.7&-1.7&-0.7\\
		  1.3& 2.3&-1.7&-0.7& 0.3
		\end{pmatrix},\\
		\lambda_k&=1/5\text{ for }k=1,\dots,5,\\
		g_{\min}&=1.22, \text{ and }
		\langle\hat L\rangle=1.03.
	\end{aligned}
	\end{align}
	For the tripartition $\{1\}:\{2,3\}:\{4\}$, we have
	\begin{align}
	\begin{aligned}
		\{\alpha^{(j)}_k\}&=
		\begin{pmatrix}
		  -0.7& 0.3& 1.3& 2.3\\
		  -2.0&-1.0&0.0&1.0\\
		  -2.0&-1.0&0.0&1.0\\
		  0.7&-2.3&-1.3&-0.3
		\end{pmatrix},\\
		\lambda_k&=1/4\text{ for }k=1,\dots,4,\\
		g_{\min}&=0.332, \text{ and }
		\langle\hat L\rangle=0.284.
	\end{aligned}
	\end{align}
	For the bipartition $\{1,2\}:\{3,4\}$, we have
	\begin{align}
	\begin{aligned}
		\{\alpha^{(j)}_k\}&=
		\begin{pmatrix}
		  -0.7& 0.3& 1.3\\
		  -0.7& 0.3& 1.3\\
		  0.7&-1.3&-0.3\\
		  0.7&-1.3&-0.3
		\end{pmatrix},\\
		\lambda_k&=1/3\text{ for }k=1,\dots,3,\\
		g_{\min}&=0.167, \text{ and }
		\langle\hat L\rangle=0.132.
	\end{aligned}
	\end{align}
	For the bipartition $\{1\}:\{2,3,4\}$, we have
	\begin{align}
	\begin{aligned}
		\{\alpha^{(j)}_k\}&=
		\begin{pmatrix}
		  -0.7& 0.3& 1.3\\
		  0.7&-1.3&-0.3\\
		  0.7&-1.3&-0.3\\
		  0.7&-1.3&-0.3
		\end{pmatrix},\\
		\lambda_k&=1/3\text{ for }k=1,\dots,3,\\
		g_{\min}&=0.167, \text{ and }
		\langle\hat L\rangle=0.132.
	\end{aligned}
	\end{align}

\end{document}